\documentclass{aastex631} 

\graphicspath{{./}{figures/}}
\shortauthors{Chakrabarty et al.}
\usepackage{xcolor}
\usepackage{soul}
\usepackage{hyperref}
\usepackage{xcolor}

\begin{document}

\title{Energetic ($<$ 2 MeV) ion fluxes measured by ASPEX-STEPS on board Aditya-L1 during its earth-bound phase} 

\author{Dibyendu Chakrabarty}
\affiliation{Physical Research Laboratory, Ahmedabad - 380009, India}

\author{Bijoy Dalal}
\affiliation{Physical Research Laboratory, Ahmedabad - 380009, India}

\author{Santosh Vadawale}
\affiliation{Physical Research Laboratory, Ahmedabad - 380009, India}

\author{Aveek Sarkar}
\affiliation{Physical Research Laboratory, Ahmedabad - 380009, India}

\author{Shiv Kumar Goyal}
\affiliation{Physical Research Laboratory, Ahmedabad - 380009, India}

\author{Jacob Sebastian}
\affiliation{Physical Research Laboratory, Ahmedabad - 380009, India}

\author{Anil Bhardwaj}
\affiliation{Physical Research Laboratory, Ahmedabad - 380009, India}

\author{P. Janardhan}
\affiliation{Physical Research Laboratory, Ahmedabad - 380009, India}

\author{M. Shanmugam}
\affiliation{Physical Research Laboratory, Ahmedabad - 380009, India}

\author{Neeraj Kumar Tiwari}
\affiliation{Physical Research Laboratory, Ahmedabad - 380009, India}

\author{Aaditya Sarda}
\affiliation{Physical Research Laboratory, Ahmedabad - 380009, India}

\author{Piyush Sharma}
\affiliation{Physical Research Laboratory, Ahmedabad - 380009, India}

\author{Aakash Gupta}
\affiliation{Physical Research Laboratory, Ahmedabad - 380009, India}

\author{Prashant Kumar}
\affiliation{Physical Research Laboratory, Ahmedabad - 380009, India}

\author{Manan S. Shah}
\affiliation{Physical Research Laboratory, Ahmedabad - 380009, India}

\author{Bhas Bapat}
\affiliation{Indian Institute of Science Education and Research, Pune 411008, India}

\author{Pranav R Adhyaru}
\affiliation{Physical Research Laboratory, Ahmedabad - 380009, India}

\author{Arpit R. Patel}
\affiliation{Physical Research Laboratory, Ahmedabad - 380009, India}

\author{Hitesh Kumar Adalja}
\affiliation{Physical Research Laboratory, Ahmedabad - 380009, India}

\author{Abhishek Kumar}
\affiliation{Physical Research Laboratory, Ahmedabad - 380009, India}

\author{Tinkal Ladiya}
\affiliation{Physical Research Laboratory, Ahmedabad - 380009, India}

\author{Sushil Kumar}
\affiliation{Physical Research Laboratory, Ahmedabad - 380009, India}

\author{Nishant Singh}
\affiliation{Physical Research Laboratory, Ahmedabad - 380009, India}

\author{Deepak Kumar Painkra}
\affiliation{Physical Research Laboratory, Ahmedabad - 380009, India}

\author{Abhishek J. Verma}
\affiliation{Physical Research Laboratory, Ahmedabad - 380009, India}

\author{Nandita Srivastava}
\affiliation{Udaipur Solar Observatory, Physical Research Laboratory, Udaipur - 313001, India}

\author{Swaroop Banerjee}
\affiliation{Physical Research Laboratory, Ahmedabad - 380009, India}

\author{K. P. Subramanian}
\affiliation{Physical Research Laboratory, Ahmedabad - 380009, India}

\author{M. B. Dadhania}
\affiliation{Physical Research Laboratory, Ahmedabad - 380009, India}

\begin{abstract}
During its earth-bound phase of the Aditya-L1 spacecraft of India, the Supra-Thermal and Energetic Particle Spectrometer (STEPS) of the Aditya Solar wind Particle EXperiment (ASPEX) was operated whenever the orbit was above 52000 km during 11 -- 19 September 2023. This phase of operation provided measurements of energetic ions (with energies 0.1--2 MeV) in the magnetosphere, magnetosheath, and interplanetary medium. Three interplanetary coronal mass ejections (ICME) hit the magnetosphere during this period. This provided opportunity to examine  the relative roles of ICME-generated solar energetic particles (SEPs) and substorm generated energetic ions on the magnetosphere. We approach this objective by detailed spectral analyses of energetic ion fluxes measured by two units of ASPEX-STEPS. We identify three distinctly different conditions of the north-south component of the interplanetary magnetic field (IMF $B_z$ = 0, $>$ 0, and $<$ 0) and use the derived spectral indices to understand this relative role. By combining these with the simultaneous energetic ion flux variations from the Advanced Composition Explorer (ACE) around the Sun-Earth first Lagrangian (L1) point and the Geostationary Operational Environmental Satellite (GOES) in the Earth's magnetosphere, we show that the polarity of IMF $B_z$ influences the energetic ion spectra in the magnetosphere  by modulating the interplay of the ICME-generated SEP with the energetic particles generated inside the magnetosphere by substorms. Interestingly, ASPEX-STEPS observations also indicate towards directional anisotropy based on spectral indices. This suggests spatially inhomogeneous mixing of energetic ions coming from different source processes. 
\end{abstract}

\keywords{Solar energetic particles (1491)  --- Solar wind (1534) --- Solar coronal mass ejections (310) -- Interplanetary magnetic fields (824)}

\section{Introduction} \label{sec:intro}
During both coronal mass ejection (CME) and their interplanetary counterpart (i.e., ICME) events, the terrestrial magnetosphere is hit by the solar energetic particles (SEPs). Substorms also generate energetic particles inside the magnetosphere. CMEs/ICMEs and substorms energize particles through different processes and up to different energies. Collisionless shocks are responsible for energization of particles in case of CMEs/ICMEs (e.g., \citealp{Desai_and_Giacalone_2016}) and the shock acceleration can generate particles to very high energies (sometimes, a few hundreds of MeVs). However, particles are energized in the terrestrial plasma sheet by processes like betatron, Fermi acceleration etc. \citep{Kronberg_et_al_2021}, by magnetic reconnection in the magnetotail (e.g., \citealp{Imada_et_al_2015}), and by the short-lived electric field induced due to change in magnetic field topology when the magnetotail snaps back from a tail-like configuration to a more dipolar configuration (also known as dipolarization) during geomagnetic substorms (e.g., \citealp{Wygant_et_al_1998}). However, unlike CME/ICME-shock acceleration where particles can be energized to very high energies, the magnetospheric processes energize particles up to a few hundreds of keVs. In addition, depending on the polarity of the north-south component of the interplanetary magnetic field (IMF B$_z$) in the magnetic cloud and sheath structures, ICMEs can cause geomagnetic storms (e.g., \citealp{Denton_et_al_2006}). On the contrary, changes in solar wind parameters (like polarity reversals in IMF B$_z$, changes in ram pressure etc.) in ICMEs can trigger substorms but, once triggered, substorms are not driven by solar wind drivers (e.g. \citealp{Kamide_et_al_1998}). In fact, \cite{Kamide_et_al_1998} argues that one can see lack of correlation between the variations in the disturbance storm time (Dst) index (proxy for storm) during geomagnetic storm and energetic particle injection activity that is seen at the geosynchronous orbit (tell-tale signature of substorms) which is observed during magnetospheric substorms. Therefore, determining the relative roles of ICME shocks and magnetospheric processes in controlling the energetic ion fluxes in the magnetosphere remains a challenging job till date. In this work, by “energetic ion fluxes” we refer to the fluxes of ions with energies $<$ 2 MeV. 

The entry of SEPs with energies $>$ 2 MeV received greater attention in the past owing to the relatively higher energies (e.g., \citealp{Kalegaev_et_al_2018, Filwett_et_al_2020}). The entry of $<$ 2 MeV ions into the magnetosphere did not receive adequate attention. Although $<$ 2 MeV ions are also SEPs, we use the term “energetic ions” at many places in this work to accommodate the substorm generated energetic ions that are not SEPs. These ions with energies $<$ 2 MeV are important to be monitored because of their harmful effects on space assets owing to the relative larger flux compared to those associated with higher energy SEPs. For example, decrease in power outputs of the uncovered solar cells due to exposure of these cells to low energy ($<$ 1 MeV) proton fluxes was reported by \cite{Statler_and_Curtin_1971}. Astrophysical and cosmological observations are hampered if these protons enter inside X-ray telescopes \citep{O’Dell_et_al_2000, Walsh_et_al_2014, Fioretti_et_al_2016}. \cite{O’Dell_et_al_2000} also reported damage of the front-illuminated charge-coupled devices (CCDs) in the Advanced CCD Imaging Spectrometer (ACIS) on board the Chandra X-ray Observatory by 100--300 keV protons. 

It is known that energetic ions ($>$ 0.1 MeV) can infiltrate into the magnetosphere through the field lines (mostly connected with the polar caps of the Earth) that open up after reconnecting with the interplanetary magnetic field (IMF) lines in the dayside or through near-equatorial dayside magnetopause via slow diffusion process \citep{Tverskoi_et_al_1973, Paulikas_1974, Scholer_1975, Kalegaev_et_al_2018}. Based on the global magnetohydrodynamic (MHD) simulations, \cite{Richard_et_al_2002} showed that protons with energies less than 10 MeV can enter the magnetosphere along open field lines and more energetic particles can directly penetrate the dayside magnetopause. Ions with relatively low energy ($\approx$ 1 MeV) can reach the last closed field line of the polar cap of Earth and can penetrate into closed field line regions through gradient and curvature drifts \citep{Scholer_1975}. Radial diffusion of these particles can also occur due to pitch-angle scattering and shell splitting of the magnetospheric magnetic field lines. \cite{Kress_et_al_2005} presented prompt trapping of solar energetic particles in the inner magnetosphere by sudden compression of the magnetosphere. Earth’s bow shock can also accelerate ``diffuse'' ions from the incoming solar wind by coupled hydromagnetic waves (e.g., \citealp{Lee_1982}). Some of these accelerated ions can propagate downstream of the bow shock in the magnetosheath region. Observations by \cite{Zong_et_al_1999} even suggested leakage of ring current ions into the magnetosheath region during the recovery phase of geomagnetic storms. 

\cite{Farrugia_et_al_1993} reported simultaneous observations of $>$ 1 MeV protons at the L1 point and the north lobe of the Earth’s magnetotail region during the passage of magnetic clouds (ICMEs). These authors suggested that long-lasting negative IMF $B_z$ provided access to these particles to the Earth’s magnetosphere. \cite{Richard_et_al_2002} brought out the importance of the southward IMF condition for a more favorable injection of SEPs into the magnetosphere. Most of the earlier studies focused on entry of highly energetic ions into the magnetosphere during geomagnetic storms and recovery phase of the storms. As energetic ions associated with an ICME can reach the Earth’s location well before the physical arrival of the ICME structure, the IMF $B_z$ condition before the arrival of the ICME can regulate the entry of $<$ 2 MeV energetic ions into the magnetosheath and magnetosphere regions. Further, studies making attempts to delineate the influence of SEPs from the substorm generated energetic ions with energies $<$ 2 MeV ions on the magnetospheric energetic particle fluxes are, to the best of our knowledge, sparse. Hence, measurements of the energetic ions in the terrestrial magnetosphere by the energetic particle spectrometer on board Aditya-L1 (AL1) mission (e.g.,\citealp{Tripathi_et_al_2022}) of India during the earth-bound phase is relevant and important. 

AL1 is the first dedicated, Indian observatory class solar mission that was launched at 11:50 Indian Standard Time (IST, UTC + 5.5 hr) on 02 September 2023 from India’s spaceport, Sriharikota. The satellite houses four remote sensing and three in-situ experiments. Aditya Solar wind Particle EXperiment (ASPEX) is one of the three in situ experiments, which comprises of two particle spectrometers –- Solar Wind Particle Spectrometer (SWIS) and Supra-Thermal and Energetic Particle Spectrometer (STEPS). ASPEX-SWIS measures bulk solar wind and ASPEX-STEPS measures suprathermal and energetic particles. The details of these spectrometers are available in \cite{Kumar_et_al_2025} and \cite{Goyal_et_al_2025}, respectively. ASPEX-STEPS was the first instrument to be switched on (after a week of launch) on 10 September 2023 and was operated during the earth-bound orbits (six closed orbits around the Earth) during 11--19 September, 2023 whenever the spacecraft was above 52000 km ($\approx$ $8R_E$). This was done to ensure that ASPEX-STEPS measured the energetic ions outside the earth’s radiation belt and ring current region. The work of Saikin et al. (2021) suggests that the outer radiation belt is extended up to $\approx 7 R_E$. Further, \cite{Wang_et_al_2024} suggests that the ring current region can extend up to $\approx 7 R_E$. Although unambiguous confirmation of the extension of the outer radiation belt and ring current region during the present observation period is not possible through the present set of measurements, it is assumed in this work that AL1 made the measurements outside the Earth’s radiation belt and ring current region based on the radial variability of the outer boundaries of these regions reported in the literature as mentioned. Several earth-bound orbits were completed before the satellite was injected into the trans-L1 orbit (cruise phase) on 19 September 2023. Out of the six detector units of ASPEX-STEPS, which point at six different directions, two units -- named as Parker Spiral (PS) and North Pointing (NP) -- made measurements of energetic ions in the terrestrial magnetosphere, magnetosheath, and in the interplanetary (IP) medium. Fortuitously, during these earth-bound orbits, three ICMEs hit the magnetosphere and three different conditions of the north-south component of the interplanetary magnetic field (IMF $B_z$ $\approx$ 0, $>$ 0 or northward and $<$ 0 or southward) were met. Further, a number of magnetospheric substorms also occurred during this period. In this work, we evaluate the energetic ion fluxes of the magnetosphere due to the interplay of energetic ions coming from the ICME shocks and substorms during three IMF B$_z$ conditions. 

In the ensuing section, we present a brief overview of the datasets used in this paper. Section 3 describes the detailed methodology, observations, and analysis. We present the results in Section 4. We discuss the results in Section 5, which is followed by conclusions in Section 6.  

\section{Datasets and cross-validation}
STEPS sub-system of ASPEX on board the Aditya-L1 (i.e., ASPEX-STEPS) is designed to measure energetic ions from six different directions with the help of six detector/sensor units - Sun Radial (SR), Parker Spiral (PS), InterMediate (IM), Earth Pointed (EP), North Pointed (NP), and South Pointed (SP). The details of ASPEX-STEPS can be found in \cite{Goyal_et_al_2018, Goyal_et_al_2025}. During the earth-bound phase, two ASPEX-STEPS units (PS and NP) were kept operational and these units provided species (primarily $H^+$ and $He^{2+}$) integrated ion flux below $\approx$ 6 MeV per nucleon. In this study, we use species-integrated ion flux (with energies $<$ 2 MeV) obtained from these two detector units. As discussed in \cite{Goyal_et_al_2018, Goyal_et_al_2025}, NP is a single window detector unit with dead layer thickness of 0.2 $\mu$m. On the other hand, PS unit comprises of a stack of custom-designed dual window Si-Pin detector and stacked scintillator plus silicon photomultiplier (SiPM) detector assembly. The dual window Si-PIN detector has two active regions that have different dead layer thickness - 0.1 $\mu$m thickness for the circular inner detector (PS-Inn the diameter of which is 7 mm) and 0.8 $\mu$m thickness for the annular (PS-Out, 7 – 18 mm diameter) outer detector. Because of very thin dead layer thickness of the PS-Inn detector, we observe some interesting spectral signatures corresponding to multiple species. The deconvolution of the observed spectral signatures in PS-Inn detector is a work in progress and hence, species integrated energetic ion flux from the PS-Outer detector has been utilized in this work. Therefore, subsequently, PS unit refers to PS-Out detector only. Another important point to note that energy of incident ions is measured in both PS and NP detectors through 256 analog to digital converter (ADC) channels. Here we have used ion fluxes from the low energy, high gain (0-127) ADC channels to cover the energy range (0.1 – 2 MeV) under consideration.

In addition to ASPEX-STEPS data, we use ion-flux data from the Electron Proton and Alpha Monitor (EPAM, \citealp{Gold_et_al_1998}) on board the Advanced Composition Explorer (ACE, \citealp{Stone_et_al_1998}) satellite at the L1 point. Ion fluxes (with energies in the range of 0.07--1.89 MeV) from the Low-Energy Magnetic Spectrometer (LEMS120, \citealp{Gold_et_al_1998}), which is one of the two magnetic telescopes of EPAM and oriented at an angle of $120^o$ with the spin axis (pointing towards the Sun) of ACE, are used. Hereafter, we use EPAM-LEMS120 to refer to this instrument. In addition, we also use the magnetospheric proton and electron fluxes obtained from the Magnetospheric Particle Sensor – High (MPSH), a subsystem of the Space Environment In-Situ Suite (SEISS, \citealp{Galica_et_al_2016}) on board the Geostationary Operational Environmental Satellite (GOES) – 18. Hereafter, SEISS-MPSH refers to this instrument on board GOES-18. GOES-18 orbits the Earth at an altitude of around $6.6R_E$. Although ASPEX-STEPS measurements have been compared and validated with respect to EPAM-LEMS120 in our earlier works \citep{Goyal_et_al_2025, Sebastian_et_al_2025} and the article of Sebastian et al. (2025) in GSICS Newsletter (Vol. 18, No. 4, 2025, doi: 10.25923/gmzc-9a28), we repeat this exercise (Figures \ref{fig:A1} and \ref{fig:A2} in the Appendix section here for another interval for the sake of completeness. Further, electron fluxes measured by SEISS-MPSH in the energy range 0.07 – 0.45 MeV have also been utilized to check occurrence of substorms. It is to be noted that SEISS-MPSH data are averaged over all of its five telescopes. We also compare the proton fluxes from SEISS-MPSH with the ion fluxes obtained from ASPEX-STEPS while it was in the magnetosphere. The comparison of temporal variations and correlations are included as Figure \ref{fig:A3} and \ref{fig:A4}, respectively, in the Appendix section. 

\begin{figure}[ht!]
\plotone{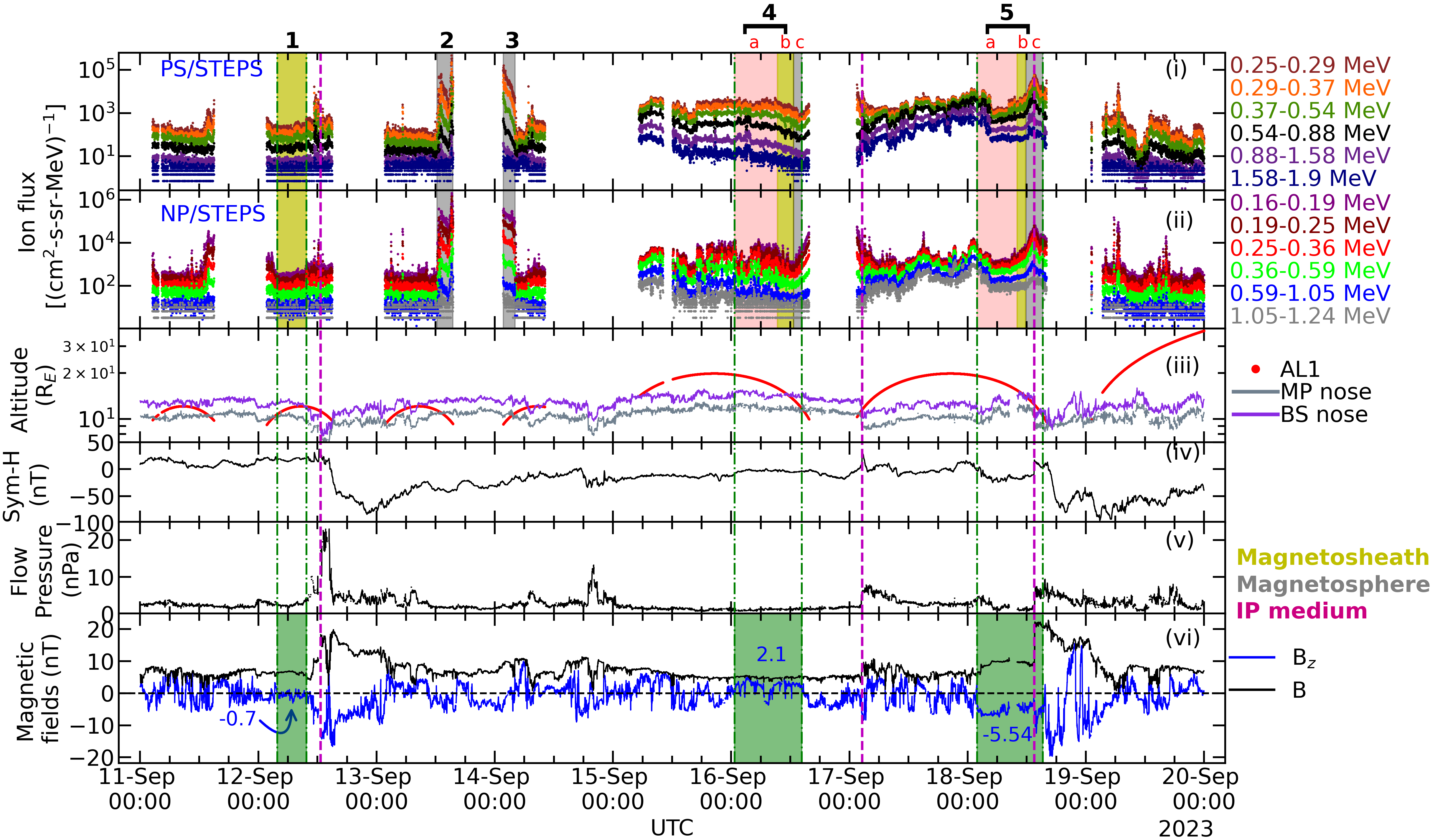}
\caption{Panels (i) and (ii) present the ion fluxes at different energy channels as measured by PS and NP units of the ASPEX-STEPS. The energy channels are mentioned at the right side of these panels. The stand-off distance of the nose of the Earth’s bow shock (violet, calculated from the position coordinates available at \url{https://cdaweb.gsfc.nasa.gov/index.html}) and magnetopause (gray, calculated using Eq. \ref{eq:Eq_1}) are plotted in panel (iii). The distance of AL1 is shown by red dots. Variations in the Sym-H are shown in panel (iv). Solar wind flow pressure is plotted in panel (v). Panel (vi) shows the variations in IMF $B_z$ (blue) and magnitude of IMF (black). The magenta colored vertical dashed lines represent arrival times of the shocks associated with three ICMEs at L1 point. The green shaded intervals in panel (vi) from left to right show different IMF $B_z$ (IMF $B_z$ $\approx$ 0, $>$ 0, and $<$ 0, respectively) conditions. The average value of IMF $B_z$ during these intervals are mentioned in blue. The yellow, pink, and gray shaded intervals in panels (i) and (ii) refer to intervals when AL1 was in the magnetosheath, interplanetary medium, and inside the magnetosphere, respectively. \label{fig1}}
\end{figure}
\section{Observations and methodology}
Panels (i) and (ii) of Figure \ref{fig1} show the variations in the energetic ion fluxes during 11-19 September 2023, as measured by the ASPEX-STEPS-PS and ASPEX-STEPS-NP detector units respectively. The altitude of AL1 (red), stand-off distances of bow shock nose (violet), and magnetopause nose (gray) during the observation period are shown in panel (iii) of Figure \ref{fig1}. Panels (iv), (v), and (vi) show, respectively, the variations in the Sym-H index, solar wind flow pressure, and IMF (magnitude (B) and z-component (B$_z$)) in the geocentric solar ecliptic (GSE) coordinate system. Data for solar wind parameters (e.g., solar wind flow pressure, IMF), sym-H, coordinates of the bow shock nose are taken from \url{https://cdaweb.gsfc.nasa.gov/index.html}.

During 11-19 September 2023, three ICMEs hit the earth. The vertical dashed lines in magenta shown in Figure \ref{fig1} represent the arrival times of the shocks associated with the three ICMEs at the Earth’s location. This is consistent with the sharp jumps in the solar wind flow pressure as shown in panel (v). Note, while the ICMEs on 17-18 September (ICME-2) and 18-19 September (ICME-3) are enlisted in the Richardson and Cane’s catalog (\url{https://izw1.caltech.edu/ACE/ASC/DATA/level3/icmetable2.htm#(c)}), the ICME-1 during 12-13 September is not listed in this catalog. It can be seen from panel (iv) of Figure \ref{fig1} that there are two moderate geomagnetic storms caused by ICME-1 and ICME-3 when minimum Sym-H values touched -80 nT.  Therefore, ICME-1 and ICME-3 generated nearly equivalent geomagnetic impact in terms of the peak negative amplitude of Sym-H providing an opportunity to evaluate and compare the impact on the energetic ion fluxes. ICME-2, on the other hand, did not cause any significant and consistent depression in Sym-H.

\begin{figure}[hb!]
\plotone{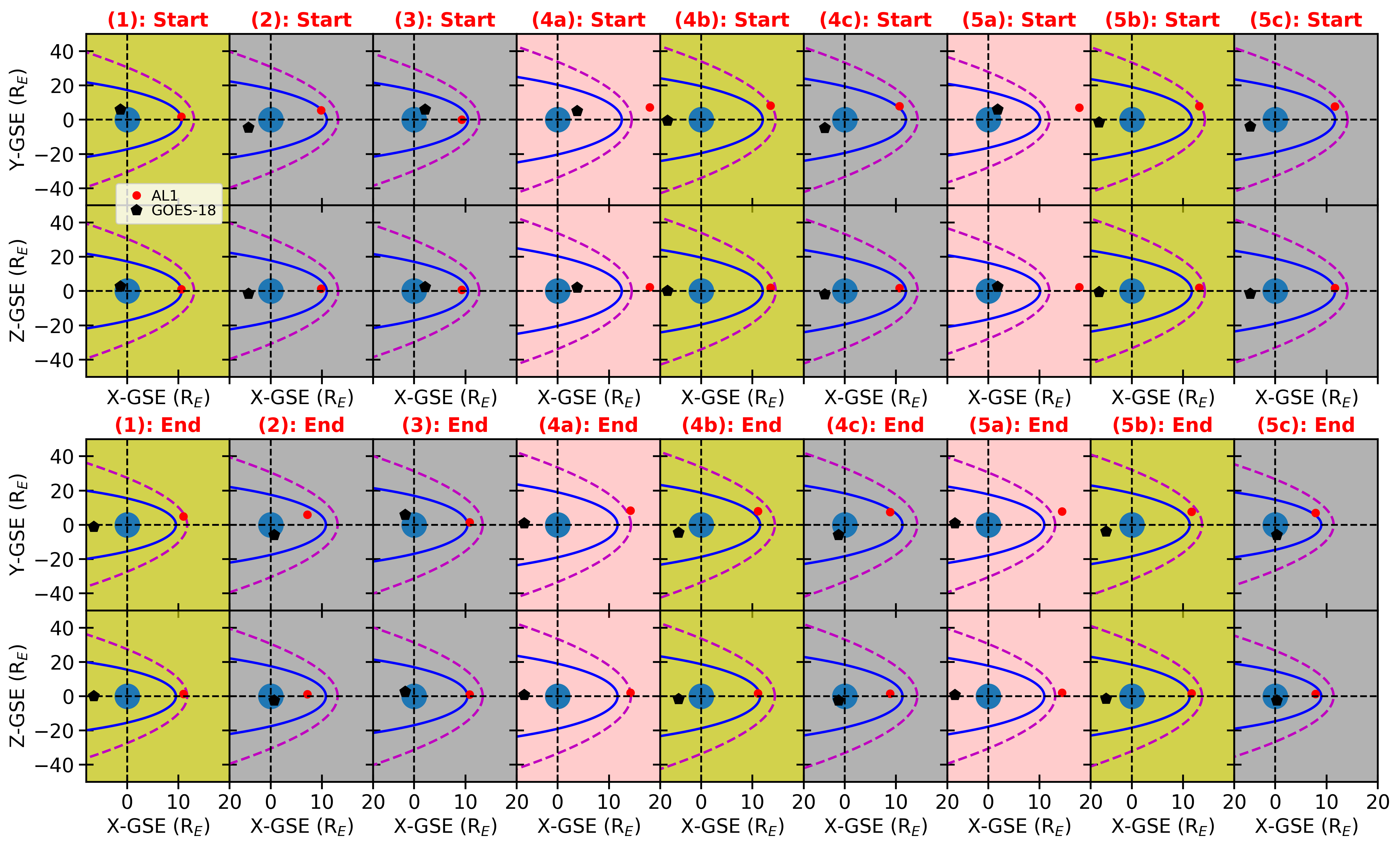}
\caption{Locations (in the XY and XZ planes of GSE coordinate system) of AL1 (red dot) and GOES-18 (black pentagon) are shown with respect to the magnetopause (blue solid curves) and bow shock boundaries (magenta dashed curves) at the start and end of all the selected intervals that are mentioned in red at the top of each pair of panels from top to bottom of each column. The distances are in terms of the Earth’s radius ($R_E$ $\approx$ 6400 km). The background color in each panel is kept same as in Figure \ref{fig1} to refer the location of the spacecraft in magnetosheath, magnetosphere, and IP medium. \label{fig2}}
\end{figure}

\subsection{Selection of intervals based on location of AL1 and IMF $B_z$ condition}
As mentioned earlier, orbits of AL1 were highly elliptical and ASPEX-STEPS detectors were switched on when AL1 was above 52000 km ($8.125 R_E$, considering $R_E$=6400 km as the radius of the Earth) from the surface of the Earth. The gaps in the AL1 altitude (panel (iii) of Figure \ref{fig1}) refer to those intervals when AL1 was below the above-mentioned height. The variations in the stand-off distances of the bow shock and magnetopause nose are directly connected to the changes in the solar wind parameters. In fact, the magnetopause nose distance is calculated, following \cite{Kivelson_and_Russell_1995}, as
\begin{equation}
    L_{mp} (R_E) = 107.4(n_{sw}u_{sw}^2)^{-\frac{1}{6}} \label{eq:Eq_1}
\end{equation}
where $n_{sw}$ and $u_{sw}$ are the proton number density and solar wind bulk speed. $n_{sw}$ and $u_{sw}$ values are taken from the CDAWeb (\url{https://cdaweb.gsfc.nasa.gov/index.html}). We present $L_{mp}$ values in panel (iii) of Figure \ref{fig1}. The positions of the bow shock nose shown in panel (iii) of Figure \ref{fig1} are available in the CDAWeb (\url{https://cdaweb.gsfc.nasa.gov/index.html}). The methodology for the calculation of the stand-off distance of the bow shock nose is given at \url{https://omniweb.gsfc.nasa.gov/html/HROdocum.html#ap4}.

We now focus on five intervals in Figure \ref{fig1} for further analysis. These are denoted as 1, 2, 3, 4, and 5. Further, the intervals 4 and 5 are divided into three (a, b, and c) additional sub-intervals. As can be seen from Figure \ref{fig1} (vi), IMF $B_z$ was nearly zero (with mean value of -0.7 nT) during interval 1, predominantly positive (northward with mean value of 2.1 nT) during interval 4, and negative (southward, with mean value of -5.54 nT) during interval 5. These intervals, selected based on IMF $B_z$ polarity, are marked by the green shaded regions in panel (vi) of Figure \ref{fig1} and these green shaded regions are sandwiched between the green dash-dot vertical lines. The yellow, pink and gray shaded intervals in panels (i) and (ii) are further sub-divisions within the green shaded intervals and these are the durations when AL1 was inside the magnetosheath, IP medium, and inside the magnetosphere, respectively. In order to find out in which region AL1 was at an interval, we have modeled the boundaries of magnetopause \citep{Shue_et_al_1997} and bow shock \citep{Chao_et_al_2002}. The magnetopause boundary is modeled using the following formula as discussed in \cite{Shue_et_al_1997}.  
\begin{equation}
    r = r_0\left(\frac{2}{1 + cos\theta}\right)^\alpha \label{eq:Eq_2}
\end{equation}
where $r_0$, $\alpha$, and $\theta$ are the stand-off distance, flaring angle, and the angle between the Earth-Sun line and the direction of r, respectively. The shape of the magnetopause in the night side is determined by $\alpha$. For $\alpha = 0.5$, the magnetopause boundary behaves asymptotically. The magnetopause is closed for $\alpha < 0.5$ and it expands with increasing distance from the Earth for $\alpha > 0.5$. We use $\alpha = 0.7$ similar to \cite{Shue_et_al_1997} to obtain a magnetopause boundary which is expanding in the night side. Since we are using STEPS data when AL1 was in the dayside only, this choice does not alter the results of this work. We feed the $L_{mp}$ (from Eq. \ref{eq:Eq_1}) values as $r_0$ in Eq. \ref{eq:Eq_2}. 

The shape of the bow shock surface is described by similar expression as in Eq. \ref{eq:Eq_2} (from \citealp{Chao_et_al_2002}), 
\begin{equation}
    r = r_0\left(\frac{1 + \epsilon}{1 + \epsilon cos\theta}\right)^\alpha \label{eq:Eq_3}
\end{equation}
where $\epsilon$ is an eccentricity factor and $\alpha$ is the tail flaring parameter. We calculate the bow shock nose distance from the position coordinates (x, y, z) in GSE coordinate system. The bow shock nose is related to r as $r = \sqrt{(x^2 + y^2 + z^2)}$. We feed these $r$ values in Eq. \ref{eq:Eq_3}. According to \cite{Chao_et_al_2002}, $\epsilon = 1.029$. We use $\alpha = 1.2$ similar to \cite{Chao_et_al_2002} in this case. It is to be noted that the bow shock and magnetopause are considered symmetric both in the Y and Z directions in the GSE coordinate system. Further, as the spacecraft is closer to the bow shock and magnetopause nose positions in both the XY and the XZ planes during this time, the choice of $\alpha$ will not significantly affect the determination of the location of the spacecraft with respect to magnetopause and bow shock boundaries. 

The position of the AL1 spacecraft (red dot) with respect to the magnetopause and bow shock boundaries is further illustrated in Figure \ref{fig2}. Here, the position of the spacecraft, when viewed in the XY and XZ planes (GSE coordinate system), at the start (top panel) and end (bottom panel) of the selected intervals (1--5) are shown. In addition to AL1, Figure \ref{fig2} shows the location of GOES-18 satellite (black pentagon) at the start and end times of the concerned intervals.

\begin{figure}[ht!]
\plotone{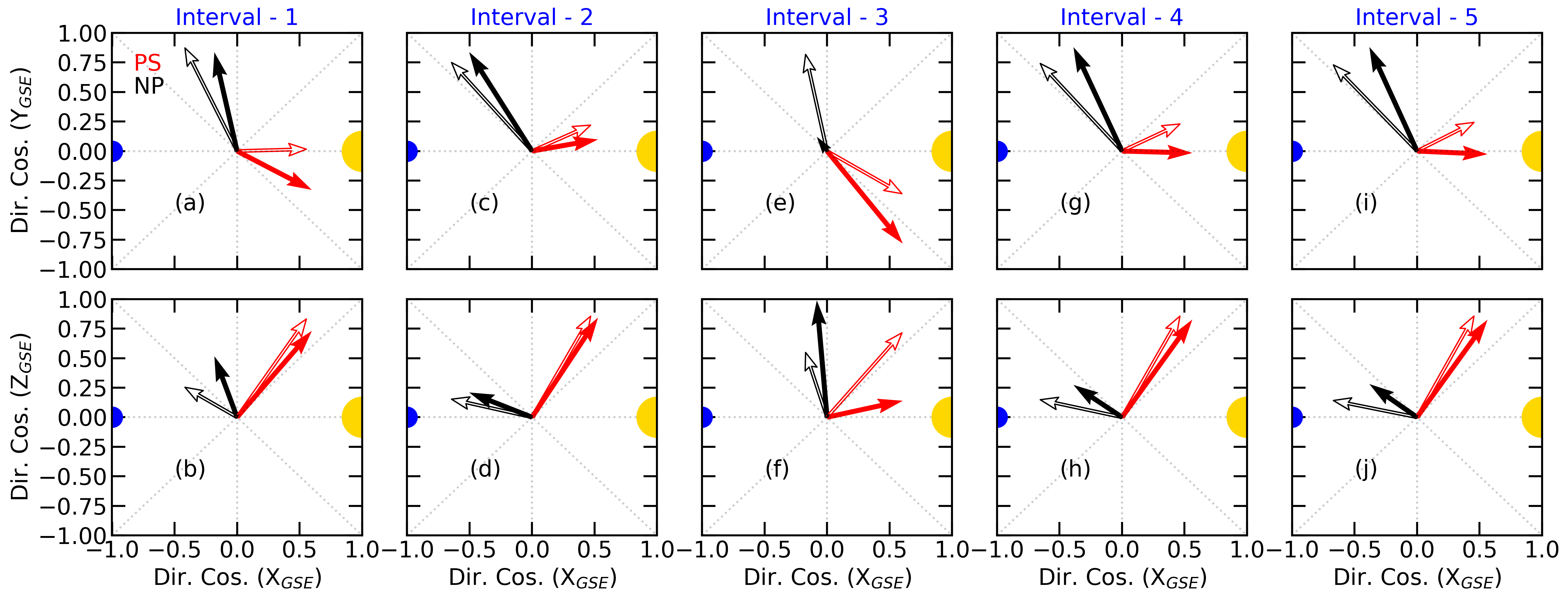}
\caption{Orientations of ASPEX-STEPS detectors with respect to GSE coordinate system during intervals 1, 2, 3, 4, and 5. The upper row presents the orientations of PS (red) and NP (black) at the start (filled arrows) of the intervals in the XY plane of GSE coordinate system. On the other hand, orientations in the XZ plane are shown in the bottom row. The unfilled arrows illustrate the orientations at the end of corresponding intervals. A rotation in the orientations of PS and NP units are observed during these intervals. Note that the length of the arrows are determined by the values of direction cosines in the corresponding planes. The Sun and Earth are represented by the yellow and blue solid circles, respectively, for reference. \label{fig3}}
\end{figure}
\subsection{Orientation of PS and NP detector units}
At this point, it is important to note that PS and NP detector units of ASPEX-STEPS were not oriented along their nominal (as planned at the L1 point) operational configurations during the observation period. Figure \ref{fig3} demonstrates the orientations of PS (red arrows) and NP (black arrows) detectors during intervals 1 (panels a, b), 2 (panels c, d), 3 (panels e, f), 4 (panels g, h), and 5 (panels i, j) with respect to GSE coordinate system. The orientations at the start and end of the intervals are shown by filled and unfilled arrows, respectively. As can be seen, PS unit was mostly looking at the Sun by making an angle with the XY-plane of the GSE coordinate system. On the other hand, NP unit was mostly oriented towards right of the Earth's dusk sector, making an angle with the GSE XY-plane. As the spacecraft was moving along its orbit, there was change in orientations of these detector units, which is illustrated in Figure \ref{fig3}.  

\begin{figure}[ht]
\centering
\includegraphics[width=0.55\textwidth, height=0.55\textwidth]{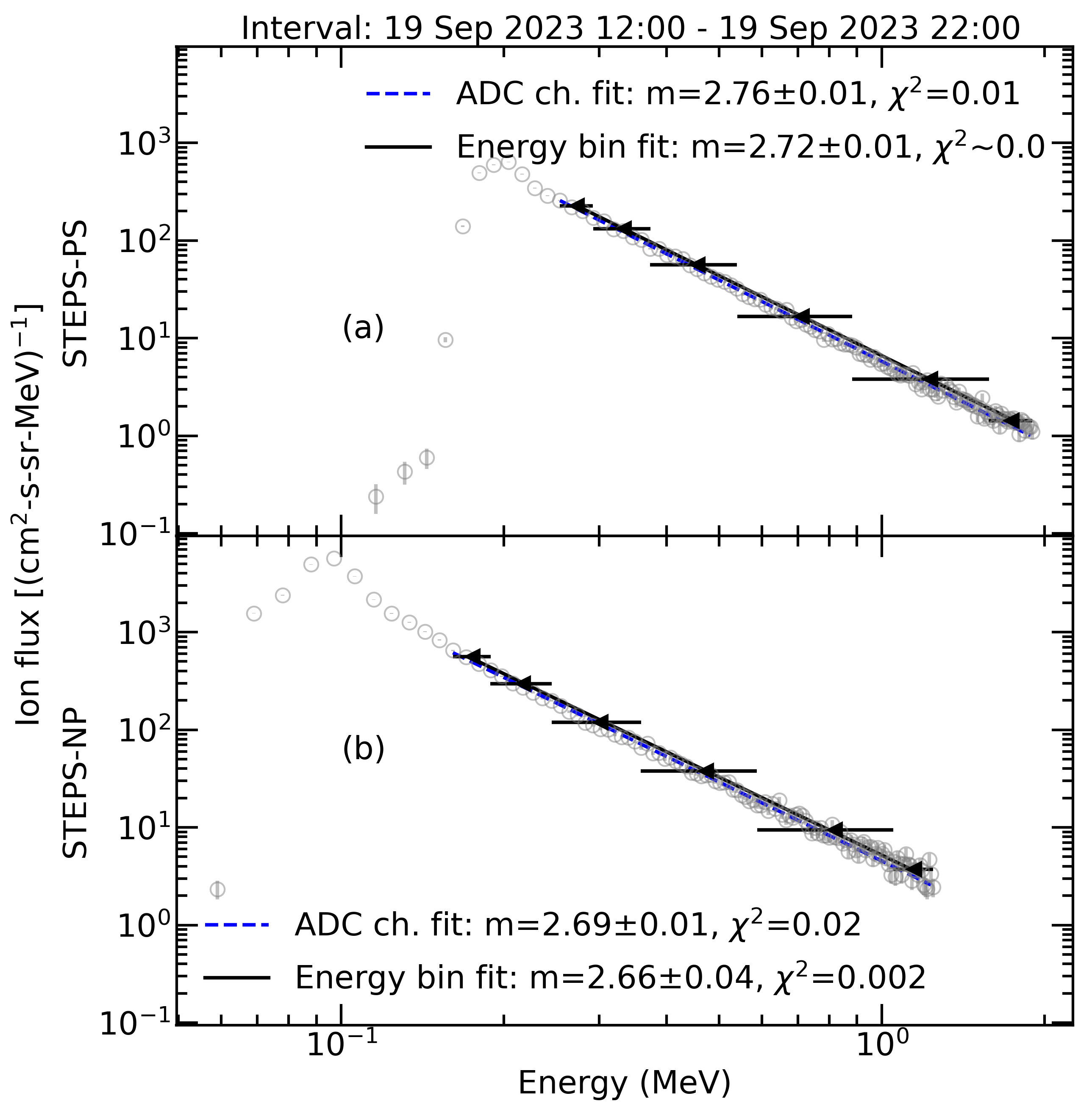}
\caption{Differential directional flux vs. energy spectrum as observed by ASPEX-STEPS-PS (panel a) and NP (panel b) on 19 September 2023 from 12:00 UT – 22:00 UT when AL1 was in the IP medium (see panel (iii) of Figure \ref{fig1}). The gray circles are the actually observed fluxes in the first 127 ADC channels of STEPS units. The vertical error bars corresponding to the gray circles represent the uncertainties in ion flux values due to counting statistics. The linear parts of these observed spectra are fitted with power law ($j \sim E^{-m}$) and are shown by blue dashed lines. The spectral index ($m$) and goodness of fit ($\chi ^2$) values are mentioned. Six energy bands are created from the linear parts of the spectra and ion fluxes corresponding to these bands are plotted by black triangles. The energy bins are also marked by the horizontal bars. The black solid fits correspond to the power law fits using the mid-points of energy bands and corresponding fluxes. The spectral indices and goodness of fits are also mentioned. Note the consistency between the spectral indices calculated from the observed spectra (ADC channel vs. Flux) and energy spectra (energy bin vs. Flux).  \label{fig4}}
\end{figure}

\subsection{Estimation of spectral index and assessment of anisotropy}
Spectral indices of ions observed by ASPEX-STEPS during the selected intervals are calculated. Figure \ref{fig4} illustrates the selection of energy range and calculation of spectral indices for PS (panel a) and NP (panel b) detector units. The raw ion spectra of incident energy vs. differential directional flux up to energy 1.89 MeV (for PS) and 1.23 MeV (for NP) detector units during 19 September 2023 12:00 - 22:00 UT are shown by gray circles in panels (a) and (b) of Figure \ref{fig4}. The energy ranges shown in this figure correspond to the first 128 ADC channels (or the high-gain channels) of these two detector units. The counting uncertainties in the flux values corresponding to these ADC channels are shown as vertical error bars. Since the vertical error bars are very small, these are not clearly visible in log-log scale. As can be seen, the nature of the spectra is different at the lower energies. This is due to low-level discriminator (LLD) threshold, which is set electronically to reduce the instrument background noise. The counts below this LLD threshold are completely zero and above this value, there is non-linear behavior up to a certain channel as seen in Figure \ref{fig4}. For a fixed LLD threshold, this channel value is always constant. We fit $J = AE^{-m}$ function (blue dashed lines) to the linear portions of the spectra (starting from 0.25 MeV for PS and from 0.16 MeV for NP) to calculate the spectral indices. The spectral indices (m) for the ADC channels vs. flux fits are 2.76 (PS) and 2.69 (NP), respectively. The reduced chi-square per data point ($\chi ^2$, defined as the sum of the square of residuals divided by the number of data points used in the fit) is also mentioned as the goodness of fit. This is a conventional approach to present the goodness of fits when the associated errors are very small (e.g., \citealp{Alderson_et_al_2023}). In the next step, we create six energy bins and fit the flux vs. energy (black solid lines) spectra. In this case, the flux values correspond to the mean energy of the energy bins. The same function (power law) has been used to fit these spectra. Counting uncertainties corresponding to the flux values have been used in all the power law fits shown in this work. The spectral indices in this case are respectively 2.72 (PS) and 2.66 (NP), which are almost the same as the spectral indices calculated earlier from the ADC vs. flux fits. We, therefore, consider the energy bin vs. flux spectra and corresponding spectral indices for interpretation throughout the paper.

Another important aspect is the assessment of anisotropy in this work. Using spectral index as a measure of anisotropy is a common practice in studies related to plasma turbulences (e.g., \citealp{wicks2010power, roberts2019anisotropy}) and for energetic ions in the magnetosheath region \citep{Kudela_et_al_1998}.  Spectral index helps to quantify the underlying stochastic process of energy transfer from higher to lower spatial scale sizes. In case of particle acceleration, spectral index is a characteristic of the underlying physical process as well. If we use only flux at a particular energy to determine anisotropic behavior, it will not indicate the energy dependence of the process involved. Therefore, spectral indices are used to quantify the directional anisotropy in the present work as this approach is physically more meaningful and provides a quantitative estimate. Further, in plasma turbulence framework, spectral indices of 1.67 and 2 are considered substantially different (e.g. see \citealp{wicks2010power, roberts2019anisotropy}). Therefore, any difference of spectral indices more than $\approx$ 0.3 can be considered quite significant. In this work, we have defined difference between spectral indices of $\approx$ 0.3 as ``mild'' and the difference of $\approx$ 0.5 is ``significant''. 

\begin{figure}[ht!]
\plotone{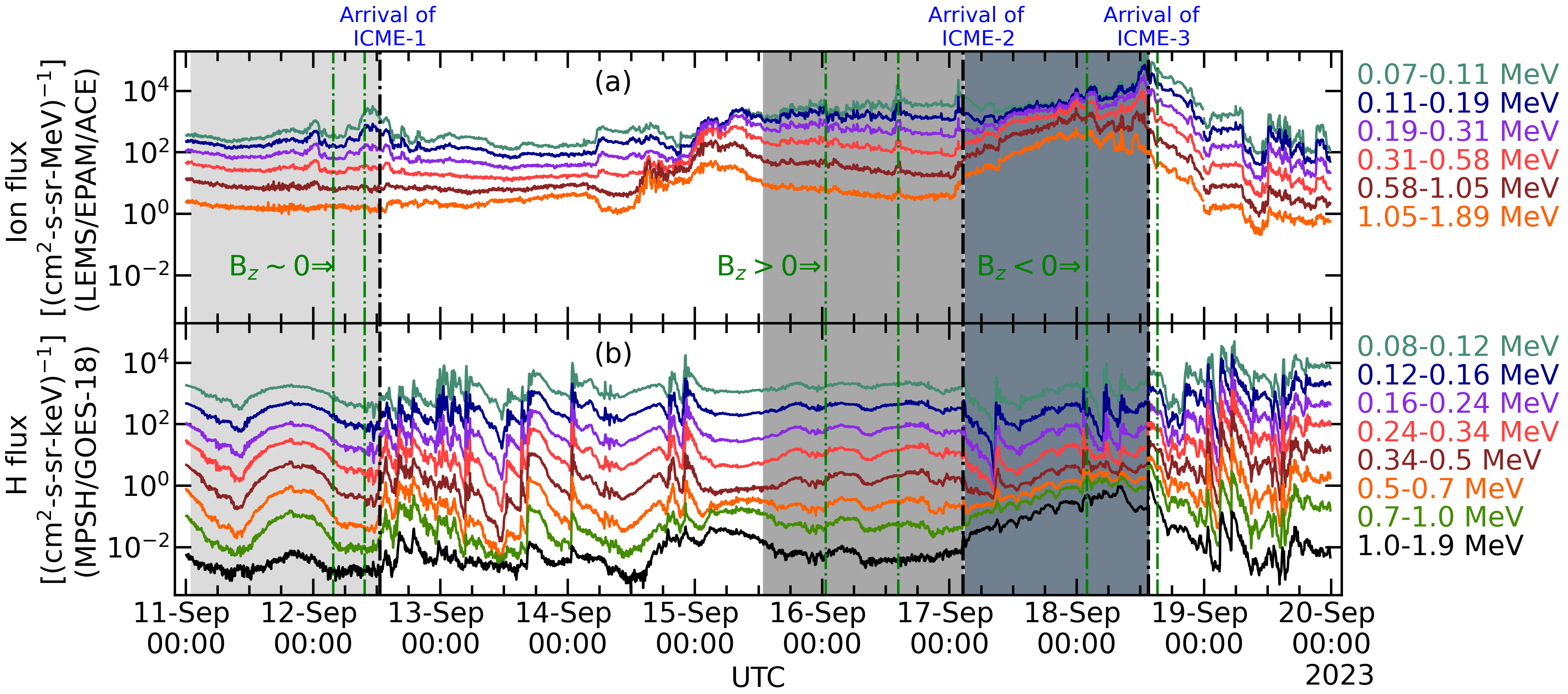}
\caption{Time series of (a) ion fluxes observed by ACE-EPAM-LEMS120 and (b) proton fluxes observed by GOES18-SEISS-MPSH at different energy channels during 11 – 19 September 2023. The arrival times of the shocks associated with ICME-1, 2, and 3 are marked by black vertical dashed lines. The extensions of intervals 1, 4, and 5 (demarcated based on IMF B$_z$, as shown in Figure \ref{fig1}) are shown by green dashed-dotted green vertical lines and corresponding IMF B$_z$ conditions are also mentioned. Fluxes during the shaded intervals (spanning 1.5 days) are used in Figure 10. \label{fig5}}
\end{figure}

\section{Results}
In this section, we focus on the variations in ion fluxes at different energy channels measured by PS and NP units of ASPEX-STEPS during intervals 1-5. It can be seen from panels (i) and (ii) of Figure \ref{fig1} that during interval 1, the ion flux levels slowly increase, whereas during interval 5, ion fluxes first drop to lower values followed by conspicuous enhancement and decrease after the shock. On the contrary, in interval 4, the flux levels slowly decrease. During intervals 2 and 3, we see sharp enhancements in the ion flux levels.

Since AL1 was sampling energetic ions from different locations (i.e., magnetosphere, magnetosheath, and IP medium as shown in Figures \ref{fig1} and \ref{fig2}) during the observation period, we take help of continuous measurements from EPAM-LEMS120 (at L1 point) and SEISS-MPSH (in the geosynchronous orbit) to assess the energetic ion fluxes in the IP medium and magnetosphere, respectively. Figure \ref{fig5} shows the temporal variations of ion and proton fluxes observed by EPAM-LEMS120 (panel a) and SEISS-MPSH (Panel b), respectively, during 11-19 September 2023. The arrival times of the concerned ICME shocks are marked by black dashed-dotted vertical lines. The intervals, selected based on IMF B$_z$ conditions (i.e., intervals 1, 4, and 5 in Figure \ref{fig1}), are also marked by green vertical dashed-dotted lines. It can be seen in panel (a) of Figure \ref{fig5} that there were multiple SEP events associated with the shocks of ICME-1, ICME-2, and ICME-3. However, the SEP variations corresponding to these ICMEs seem to be different in high energy channels. While for ICME-1, the high-energy fluxes are nearly constant, the high-energy fluxes are decreasing and increasing in the case of ICME-2 and 3, respectively. On the other hand, variations in the proton fluxes (panel (b) of Figure \ref{fig5}) show many short and long-term enhancements. The implications of the shaded intervals will be discussed in the Discussion section when we show correlation of energetic ion fluxes in the IP medium and proton fluxes inside the Earth’s magnetosphere.   

In the next part of this section,  we focus on the spectra of energetic ions measured by all the above mentioned instruments during intervals 1-5 and also derive spectral indices as demonstrated in Figure \ref{fig4}. We present these ion spectra in the following subsections based on the location of AL1 i.e., inside the magnetosphere, magnetosheath, and IP medium.

\subsection{Ion spectra when AL1 was inside the magnetosphere}
\begin{figure}[ht]
\centering
\includegraphics[width=0.65\textwidth, height=0.65\textwidth]{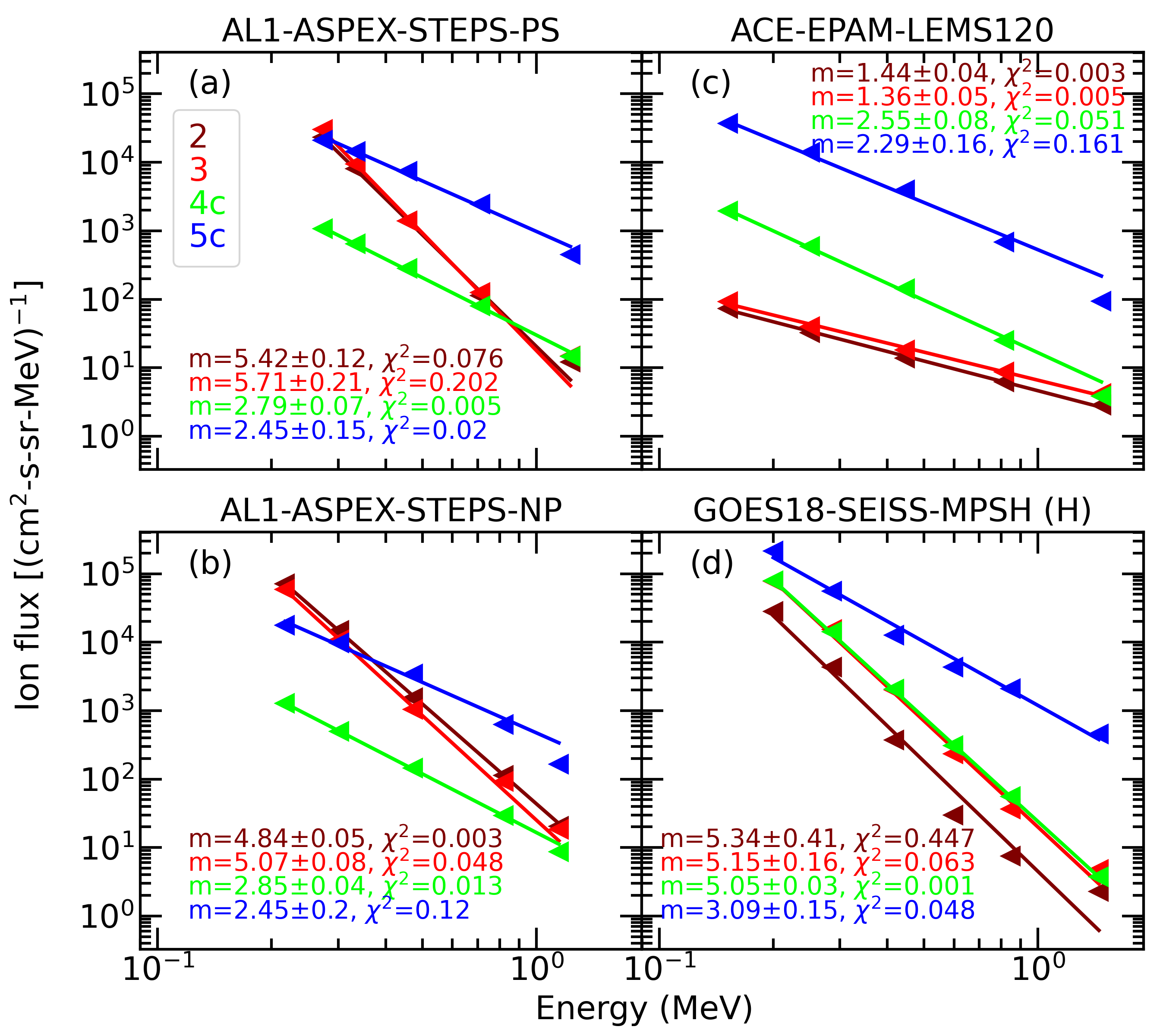}
\caption{Ion spectra observed by (a) ASPEX-STEPS-PS, (b) STEPS-NP, (c) ACE-EPAM-LEMS120, and (d) proton spectra observed by GOES18-SEISS-MPSH during intervals when AL1 was in the magnetosphere (2, 3, 4c, and 5c) are shown by different colors. The spectral indices ($m$) and $\chi ^2$ values are also mentioned. \label{fig6}}
\end{figure}

As can be seen in Figure \ref{fig1} and Figure \ref{fig2}, AL1 was inside the magnetopause boundary (i.e. in the magnetosphere region) during intervals 2, 3, 4c, and 5c. Figure \ref{fig6} presents $<$ 2 MeV ion spectra observed by ASPEX-STEPS (Panels a and b) and proton (hereafter, denoted by $H^+$) spectra observed by SEISS-MPSH (panel d) in the magnetosphere. It also shows ion spectra observed by EPAM-LEMS120 in the similar energy range during these intervals (Panel c). The spectral indices ($m$) and goodness of fits ($\chi ^2$) are mentioned for each of the spectra. The fit parameters $m$, $c (=ln (A))$, and corresponding $\chi ^2$ values are tabulated in Table \ref{tab1}.

It can be noted that in intervals 2 and 3, measurements by ASPEX-STEPS-PS (5.42 and 5.71) and ASPEX-STEPS-NP (4.84 and 5.07) units yielded spectral indices $>$ 4.5. $H^+$ spectral indices observed by SEISS-MPSH in these two intervals are closer (5.34 and 5.15) to those given by the ASPEX-STEPS detectors. On the other hand, spectral indices of energetic particles in the IP medium as observed by EPAM-LEMS120 are completely different (1.44 and 1.36) from those observed in the magnetosphere in intervals 2 and 3. We also note sharp enhancements in fluxes captured by AL1 during interval 2 and 3 similar to what we expect during substorms \citep{Chakrabarty_et_al_2008, Chakrabarty_et_al_2015}. The ion fluxes in intervals 2 and 3 are higher compared to those in 4c and comparable to those in 5c although AL1 was inside the magnetosphere in these intervals. In interval 4c, spectral indices of ions observed by ASPEX-STEPS-PS and NP detectors are 2.79 and 2.85, respectively whereas GOES H spectrum shows spectral index of 5.05 as before. This time, spectral index of ions observed by ACE-EPAM-LEMS120 is 2.55. This suggests AL1 and ACE spectral indices are closer in interval 4c while GOES spectral index is much higher (softer spectra) and nearly same as intervals 2 and 3. During interval 5c, spectral indices of ions recorded by ASPEX-STEPS units are 2.45 (PS) and 2.45 (NP). EPAM-LEMS120 recorded ion spectral index of 2.29 and SEISS-MPSH detected H spectral index of 3.09. In this case, the AL1, GOES and ACE spectral indices are not significantly different. Note, IMF $B_z$ is southward during interval 5c. Another important point can be noted here. Mild (spectral indices are different by $\approx$ 0.3) to significant (spectral indices are different by more than $\approx$ 0.5) directional anisotropies between PS and NP units are seen during intervals 2 and 3 but during intervals 4c and 5c, the directional anisotropy is either absent or negligible. 

\subsection{Ion spectra when AL1 was in magnetosheath}
\begin{figure}[ht]
\centering
\includegraphics[width=0.65\textwidth, height=0.65\textwidth]{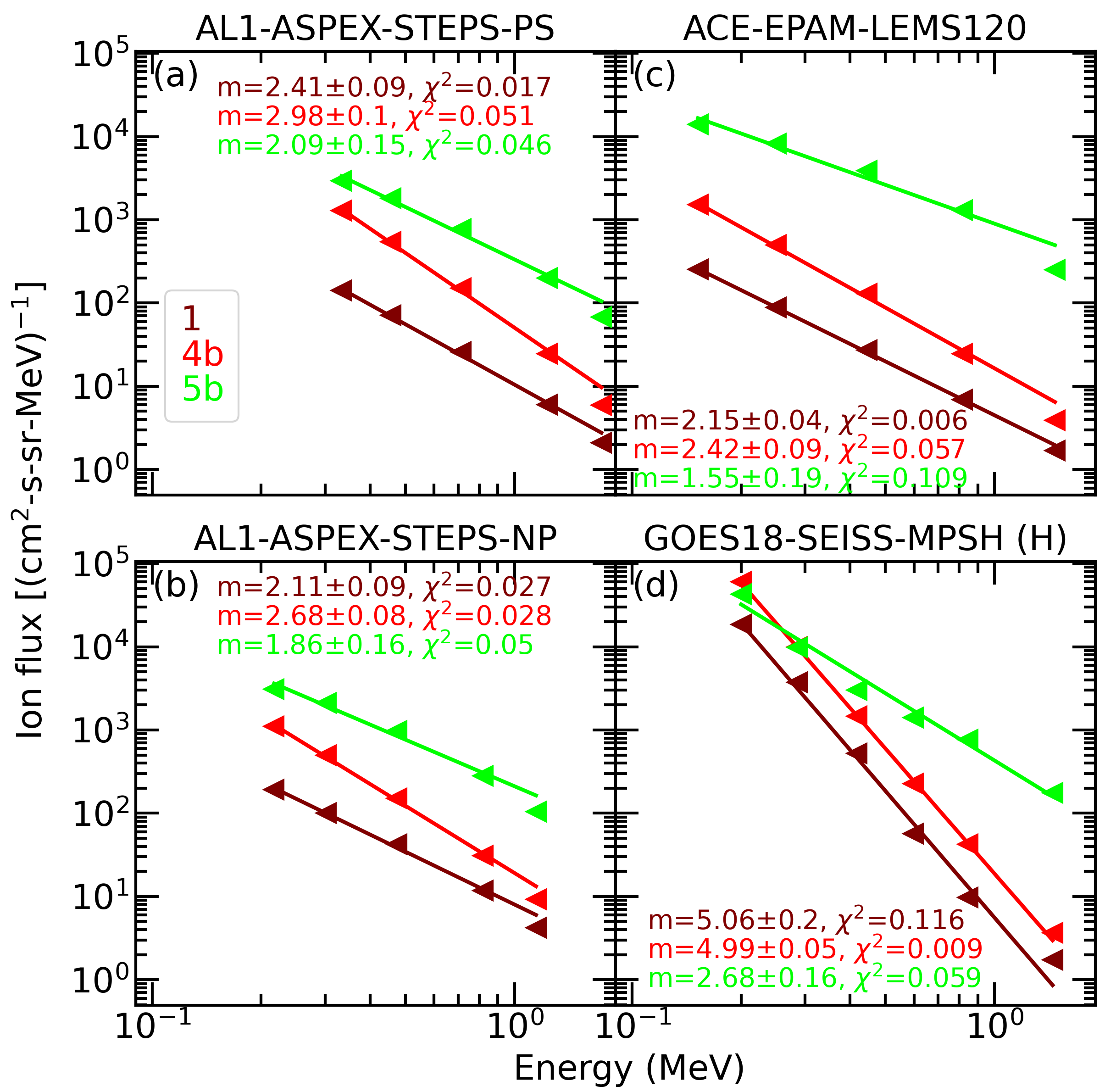}
\caption{Ion spectra observed by (a) ASPEX-STEPS-PS, (b) STEPS-NP, (c) ACE-EPAM-LEMS120, and (d) proton spectra observed by GOES18-SEISS-MPSH during intervals when AL1 was in the magnetosheath (1, 4b, and 5b) are shown by different colors. The spectral indices ($m$) and $\chi ^2$ values are mentioned. \label{fig7}}
\end{figure}

Figure \ref{fig7} illustrates the ion spectra observed by ASPEX-STEPS, EPAM-LEMS120, and H spectra observed by SEISS-MPSH during intervals 1, 4b, and 5b when AL1 was in the magnetosheath region. The fit parameters $m$, $c$, and $\chi ^2$ values are tabulated in Table \ref{tab1}. Spectral indices of ions recorded by ASPEX-STEPS detectors in the magnetosheath region are 2.41, 2.98, and 2.09 (PS) and 2.11, 2.68, and 1.86 (NP), respectively, in intervals 1, 4b, and 5b, respectively. Therefore, consistently mild (spectral indices differ by $\approx$ 0.3) directional anisotropies are noticed between the PS and NP units, while AL1 was in the magnetosheath. Energetic ions measured by EPAM-LEMS120 in the IP medium exhibited spectral indices 2.15, 2.42 and 1.55 during intervals 1, 4b, and 5b, respectively. During these intervals, SEISS-MPSH showed spectral indices 5.06, 4.99, and 2.68. Therefore, AL1 spectral indices are found to be relatively closer to those of ACE and harder than the GOES spectra during these intervals. Note, IMF $B_z$ is close to 0, predominantly northward and southward during intervals 1, 4b and 5b, respectively. 

\subsection{Ion spectra when AL1 was in the IP medium}
\begin{figure}[ht]
\centering
\includegraphics[width=0.65\textwidth, height=0.65\textwidth]{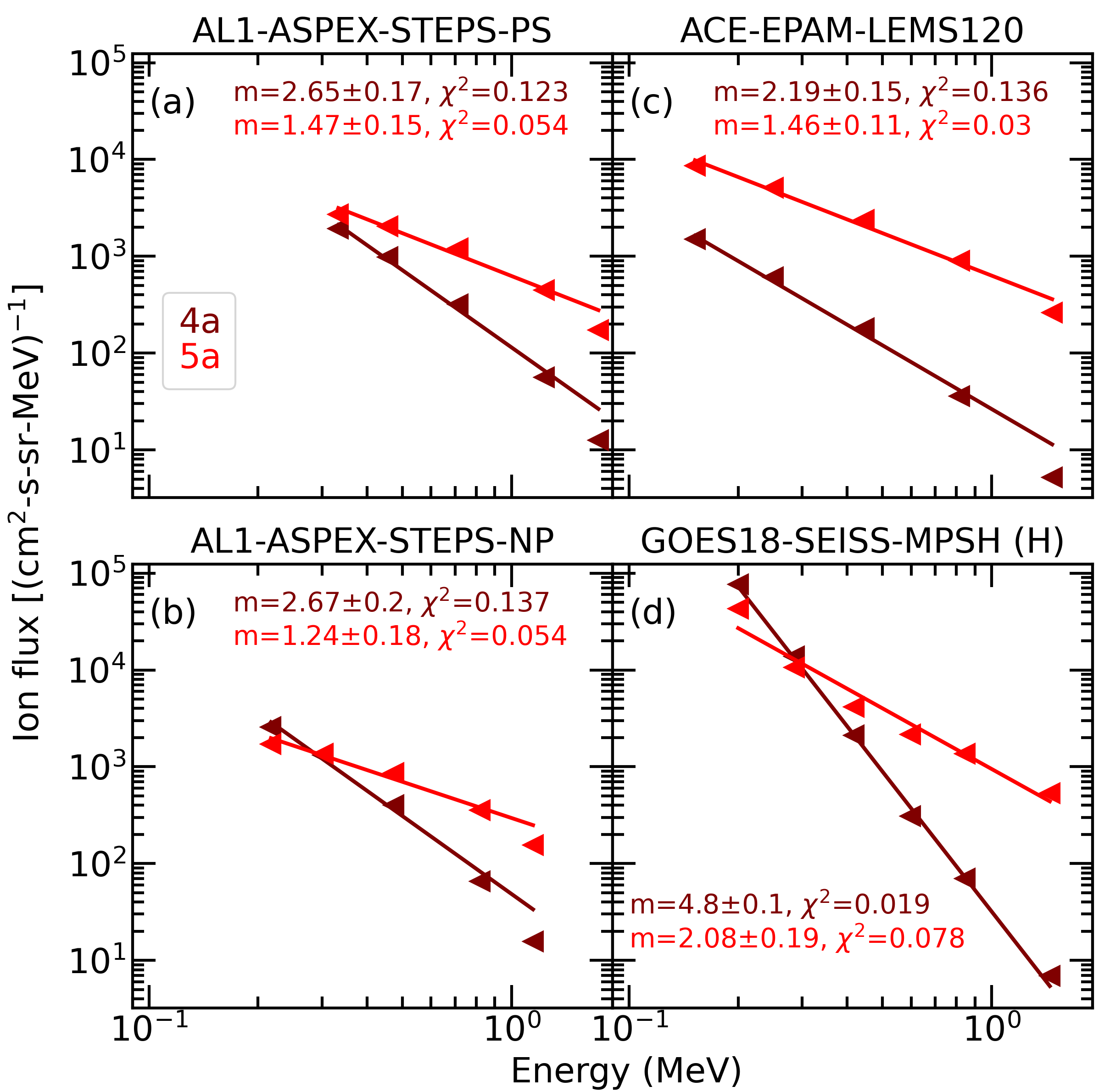}
\caption{Ion spectra observed by (a) ASPEX-STEPS-PS, (b) STEPS-NP, (c) ACE-EPAM-LEMS120, and (d) proton spectra observed by GOES18-SEISS-MPSH during intervals when AL1 was in the IP medium (4a and 5a) are shown by different colors. The spectral indices ($m$) and $\chi ^2$ values are mentioned. \label{fig8}}
\end{figure}

During the intervals 4a and 5a, AL1 measured energetic ions in the IP medium. Figure \ref{fig8} provides the spectra of ions (in the IP medium) and H (magnetosphere) in the same fashion as in Figure \ref{fig6} and Figure \ref{fig7}. As before, the fit parameters $m$, $c$, and $\chi ^2$ values are tabulated in Table \ref{tab1}. It is to be noted that the distance of AL1 from the Earth during these two intervals was $<$ $20 R_E$ (see Panel (iii) of Figure \ref{fig1}) whereas, ACE was around $244 R_E$ away from the Earth. The spectral indices of ions observed by ASPEX-STEPS detector units in intervals 4a and 5a are 2.65 and 1.47 (PS) and 2.67 and 1.24 (NP), respectively. Therefore, mild directional anisotropy is seen during interval 5a but not during interval 4a. EPAM-LEMS120 recorded spectral indices of 2.19 and 1.46 during these intervals. On the other hand, SEISS-MPSH observed H spectral indices 4.8 and 2.08, respectively, in intervals 4a and 5a. Therefore, it is seen that AL1 and ACE spectral indices are closer in both the intervals 4a and 5a compared to those of GOES. Further, hard spectra (with lower value of spectral index) are observed by all the detector units (including GOES) in interval 5a.

\begin{deluxetable*}{c|c c c|c c c|c c c|c c c}
\tablecolumns{13}
\tablenum{1}
\tablecaption{List of spectral indices and fit parameters for different intervals and locations of AL1. \label{tab1}}
\tablewidth{12pt}
\tablehead{
\colhead{} & \multicolumn{3}{c}{\textbf{ASPEX-STEPS-PS}} & \multicolumn{3}{c}{\textbf{ASPEX-STEPS-NP}} & \multicolumn{3}{c}{\textbf{EPAM-LEMS120}} & \multicolumn{3}{c}{\textbf{SEISS-MPSH}}\\
\colhead{Interval} & 
\colhead{m} & \colhead{c} & \colhead{$\chi ^2$} &
\colhead{m} & \colhead{c} & \colhead{$\chi ^2$} & 
\colhead{m} & \colhead{c} & \colhead{$\chi ^2$} & 
\colhead{m} & \colhead{c} & \colhead{$\chi ^2$}
}
\startdata
\multicolumn{13}{c}{\textbf{Magnetosphere}}\\
\hline
2 &	5.42$\pm$0.12 & -3.02$\pm$0.14 & 0.08 &	4.84$\pm$0.05 & -3.77$\pm$0.07 & 0 & 1.44$\pm$0.04 & -1.53$\pm$0.05 & 0 & 5.34$\pm$0.41 & -3.11$\pm$0.5 & 0.45 \\
3 & 5.71$\pm$0.21 & -2.88$\pm$0.25 & 0.2 & 5.07$\pm$0.08 & -3.23$\pm$0.12 & 0.05 & 1.36$\pm$0.05 & -1.88$\pm$0.07 & 0.01 & 5.15$\pm$0.16 & -4.6$\pm$0.18 & 0.06 \\
4c & 2.79$\pm$0.07 & -3.4$\pm$0.07 & 0.01 & 2.85$\pm$0.04 & -2.81$\pm$0.05 & 0.01 & 2.55$\pm$0.08 & -2.81$\pm$0.13 & 0.05 & 5.05$\pm$0.03 & -4.79$\pm$0.03 & 0 \\
5c & 2.45$\pm$0.15 & -6.89$\pm$0.13 & 0.02 & 2.45$\pm$0.2 &	-6.16$\pm$0.23 & 0.12 & 2.29$\pm$0.16 & -6.27$\pm$0.24 & 0.16 & 3.09$\pm$0.15 & -8.69$\pm$0.14 & 0.05\\
\hline
\multicolumn{13}{c}{\textbf{Magnetosheath}}\\
\hline
1 & 2.41$\pm$0.09 & -2.35$\pm$0.07 & 0.02 & 2.11$\pm$0.09 & -2.08$\pm$0.1 & 0.03 & 2.15$\pm$0.04 & -1.5$\pm$0.06 & 0.01 & 5.06$\pm$0.2 & -3.33$\pm$0.24 & 0.12 \\
4b & 2.98$\pm$0.1 & -3.92$\pm$0.08 & 0.05 & 2.68$\pm$0.08 & -2.95$\pm$0.09 & 0.03 & 2.42$\pm$0.09 & -2.8$\pm$0.14 & 0.06 & 4.99$\pm$0.05 & -4.55$\pm$0.06 & 0.01 \\
5b & 2.09$\pm$0.15 & -5.81$\pm$0.11 & 0.05 & 1.86$\pm$0.16 & -5.35$\pm$0.17 & 0.05 & 1.55$\pm$0.19 & -6.8$\pm$0.24 & 0.11 & 2.68$\pm$0.16 & -7.68$\pm$0.15 & 0.06 \\
\hline
\multicolumn{13}{c}{\textbf{IP medium}}\\
\hline
4a & 2.65$\pm$0.17 & -4.74$\pm$0.14 & 0.12 & 2.67$\pm$0.2 & -3.88$\pm$0.24 & 0.14 & 2.19$\pm$0.15 & -3.28$\pm$0.23 & 0.14 & 4.8$\pm$0.1 & -5.08$\pm$0.1 & 0.02 \\
5a & 1.47$\pm$0.15 & -6.44$\pm$0.1 & 0.05 & 1.24$\pm$0.18 & -5.69$\pm$0.18 & 0.05 & 1.46$\pm$0.11 & -6.45$\pm$0.14 & 0.03 & 2.08$\pm$0.19 & -8.46$\pm$0.16 & 0.08
\enddata
\end{deluxetable*}

\section{Discussions}
During the earth-bound phase of Aditya-L1, the spacecraft crossed the magnetosphere, magnetosheath and IP medium multiple times and measured energetic ion fluxes. In the following subsections, we group the spectral observations reported earlier based on where the spacecraft was located (like magnetosheath, magnetosphere, IP medium) and discuss the possible reasons for the variations in the energetic ion spectra due to the influence of ICME associated SEPs and substorm associated energetic ions.
\subsection{AL1 in the magnetosheath region}
Among the 9 intervals mentioned in Figure \ref{fig1}, AL1 was in the magnetosheath region during intervals 1, 4b, and 5b. Energetic ion populations in the magnetosheath may consist of ions arriving from the IP medium, bow shock (e.g., \citealp{Lee_1982, Chang_et_al_2000, Burgess_2007}), and from the Earth’s magnetosphere \citep{Scholer_et_al_1981, Sibeck_et_al_1987, Zong_and_Wilken_1999, Cohen_et_al_2016}. Although not characterized in detail, in this work, the main sources of energetic particles in the IP medium are SEPs associated with shocks driven by ICME-1, 2 and 3. According to \cite{Rodriguez_Pacheco_et_al_1998}, the typical spectral index of the energetic protons (with energies 0.36--1.6 MeV) associated with gradual SEP events in solar cycle 21 in the IP medium varies in the range 1.25-1.94. Accelerated particles from the terrestrial bow shock populate the magnetosheath region (e.g., \citealp{Lee_1982, Chang_et_al_2000}) and the typical spectral index of these particles is $\approx 4.5$ \citep{Desai_et_al_2000}. On the other hand, ions, accelerated at the magnetosphere (and leaking into magnetosheath), typically exhibit spectral indices in the range 4 -- 6 (e.g., \citealp{Fan_et_al_1975, Sarris_et_al_1976, Imada_et_al_2015}). Therefore, when energetic ions from the above-mentioned three sources mix up in the magnetosheath region, average (lying between 2 -- 5) spectral indices can be expected. In fact, \cite{Fan_et_al_1975} observed ion spectral indices of $\approx$ 3.8 in the magnetosheath region. Keeping this in mind, we evaluate the present set of observations.  

\subsubsection{Interval 1}
It can be seen in Figure \ref{fig7} and Table \ref{tab1} that during interval 1, ASPEX-STEPS-PS and NP recorded ion spectral indices of 2.41 and 2.11, respectively. While at the same time, magnetospheric $H^+$ spectrum was much softer with spectral index of 5.06 and in the IP medium, spectral index of 2.15 was observed. We see small enhancements in the energetic ions in panel (a) of Figure \ref{fig5}, which are expected to be SEPs associated with ICME-1 shock. Therefore, looking at the spectral indices it appears that during interval 1,  contribution of these SEPs dominate in the magnetosheath region. Under IMF $B_z \approx$ 0 condition, energetic ions with softer spectra (with spectral index 5.06) from the magnetosphere seem to have little contribution in the magnetosheath region. Another notable aspect is the mild anisotropy in spectral indices in ASPEX-STEPS-PS (2.41) and NP (2.11) units. Being in the magnetosheath region during interval 1, PS unit was mostly looking at the bow shock. Therefore, it is possible that bow shock accelerated energetic ions with a little softer spectra \citep{Kudela_et_al_1998, Desai_et_al_2000} contributed in mild softening (spectral index reduces by ~0.3) of PS spectra as compared to that of NP, which was oriented along the dusk sector.       

\subsubsection{Interval 4b}
During interval 4b, $H^+$ spectral index inside the magnetosphere (GOES) is 4.99 while in the IP medium (ACE) energetic ions exhibit spectral index of 2.42. AL1 recorded spectral indices of 2.98 (ASPEX-STEPS-PS) and 2.68 (ASPEX-STEPS-NP) in the magnetosheath region. Panel (a) of Figure \ref{fig5} shows a decreasing trend in the high energy ion fluxes observed by EPAM-LEMS120 during this interval and the ICME-2 shock arrived physically much later. Energetic ions from this SEP event do not have a hard spectra, which is clear from ACE measurement at that time. A decaying trend in the ion fluxes observed by ASPEX-STEPS units also can be seen in Figure \ref{fig1}. Therefore, it appears that there was a contribution of the ICME-2 shock associated ions in the magnetosheath region during (interval 4b). However, softer ion spectra observed by ASPEX-STEPS units (2.99, 2.68 for PS and NP, respectively) as compared to that (2.42) in the IP medium indicates possible additional influence of a much softer ion spectra in the magnetosheath region. Note that IMF $B_z$ was positive during interval 4b, which indicates minimal merging of IMF B$_z$ and the Earth's magnetic field in the dayside magnetopause. However, even under IMF $B_z$ $>$ 0 condition, leakage of magnetospheric ions into the magnetosheath region cannot be neglected (e.g., \citealp{Sibeck_et_al_1987}). Magnetospheric energetic ions may leak into the magnetosheath region owing to a variety of processes that include the finite gyro-radius effect for ions with energies of $\geq$ 10s of keV \citep{sorathia_et_al_2017}, Kelvin-Helmholtz instability \citep{merkin_et_al_2013, hwang_et_al_2023} for ions with energies of $>$ 30 keV, magnetopause shadowing \citep{Sibeck_et_al_1987} for ions with energies of 100s of keV, etc. In addition, bow shock accelerated ions \citep{Burgess_2007} may also contribute. A mild softer (difference of 0.3) spectra in ASPEX-STEPS-PS unit (looking at the bow shock) than NP (looking at the dusk sector), leading to a mild anisotropy, possibly hints towards this mixing. 

\subsubsection{Interval 5b} 
During interval 5b, energetic ions in the IP medium (ACE) exhibits a spectral index of 1.55. ASPEX-STEPS units in the magnetosheath region recorded spectral indices of 2.09 (PS) and 1.86 (NP). $H^+$ in the geosynchronous orbit showed a spectral index of 2.68, which is much harder as compared to the previous cases. Panel (a) of Figure \ref{fig1} shows that interval 5b corresponds to the rising phase of the SEP event associated with ICME-3 shock. Enhancements in energetic ion fluxes measured by ASPEX-STEPS units are also seen in panels (i) and (ii) of Figure \ref{fig1}. Therefore, there is clear indication of SEP influence in the magnetosheath region during interval 5b. Aided by the negative IMF $B_z$ condition, even the energetic $H^+$ ions deep inside the magnetosphere (i.e., geosynchronous orbit) also responded to these SEPs, exhibiting harder spectrum. The SEP influence in the magnetosheath region during this time is so strong that despite oriented along different directions, ASPEX-STEPS units merely recorded any directional anisotropy. However, the spectral indices (2.09 and 1.86) observed (by ASPEX-STEPS units) in the magnetosheath region are soffter than the IP medium spectral index (1.55). This is possibly due to mixture of SEPs with the energetic ions from the bow shock by the bow shock and from the magnetosphere assisted by negative IMF $B_z$ condition. 

The above discussion suggests that during intervals 4b and 5b, possible mixing of SEPs with energetic ions from the magnetosphere and bow shock took place in the magnetosheath region. On the other hand, dominant role of SEPs and bow shock modulation stand out during interval 1.   

\subsection{AL1 in magnetosphere}
AL1 was inside the magnetosphere during intervals 2, 3, 4c, and 5c. It is known that Earth’s plasma sheet acts as primary reservoir of magnetospheric energetic ions \citep{Kronberg_et_al_2021}. Ions from the plasma sheet can be energized by adiabatic processes (e.g., betatron, Fermi acceleration) when ions drift towards the Earth. Ions can also be accelerated up to $\approx$ 100 keV by quasi-steady dawn-dusk electric field. In general, high-energy (50 keV -- 1 MeV) spectra of the plasma sheet ions follow power law with exponent $\approx$ 6.5 during low to moderate auroral electrojet activity \citep{Christon_et_al_1988}. Another source of energetic ions in the magnetosphere is substorm, which are associated with magnetic field dipolarization and burst of energetic particles towards the Earth (e.g., \citealp{McPherron_1979}). Softer spectra (with spectral index 4 -- 6) are generally observed for particles accelerated during substorms. A typical example of substorm associated proton spectra is provided in Figure \ref{fig:A5}, where we calculate the spectral index of proton in the geosynchronous orbit as measured by GOES17-SEISS-MPSH during a substorm event reported in \cite{Rathi_et_al_2025}. The spectral index we get is 5.15 in that case. We have also checked GOES-18 proton spectra associated with some other substorm events on 10 January 2025, 11 January 2025, 17 January 2025, 14 February 2025, 18 March 2025, and 19 March 2025 (not shown in this work). The spectral indices in these cases vary in the range from 4.6 - 5.54. Therefore, it is apparent that substorms are often associated with spectral indices $\approx$ 5. 

\begin{figure}[ht!]
\plotone{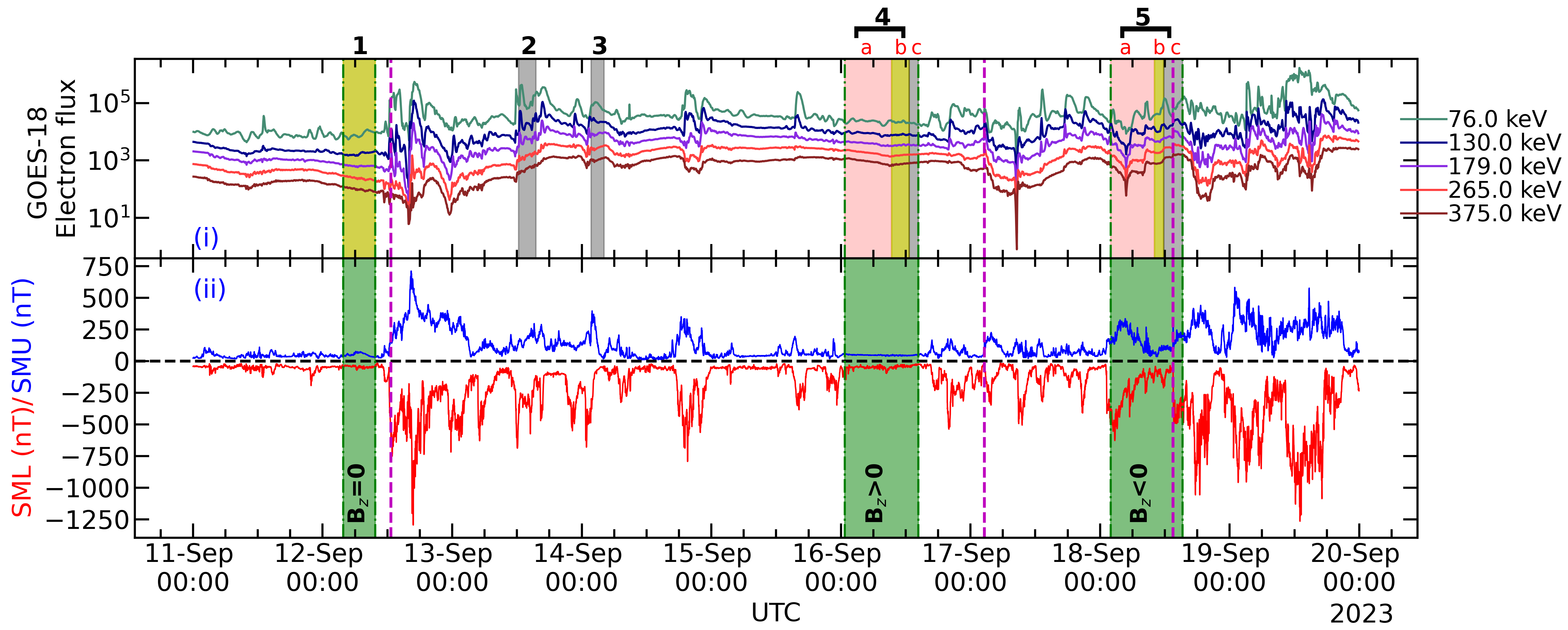}
\caption{(i) Variations of electron fluxes at four different energy channels (mentioned at the right of the top panel) during 11-19 September 2023, as observed by GOES-SEISS-MPSH. The yellow, pink, and gray shaded intervals refer to the same intervals as described in Figure \ref{fig1}, (ii) Temporal variations of SMU and SML indices (source: \url{https://supermag.jhuapl.eom paneldu/}) during the same period are shown in the bottom. The conditions of IMF $B_z$ are mentioned inside the green shaded intervals similar to Figure \ref{fig1}. The magenta vertical dashed lines indicate the arrival of the shocks/disturbances associated with three ICMEs as mentioned in Figure \ref{fig1}.  \label{fig9}}
\end{figure}

Since magnetospheric substorms can generate energetic ions (a few hundreds of keVs) inside the magnetosphere, it is important to check the presence of substorms during the entire period under consideration. In order to identify the occurrence of substorms, we evaluate electron fluxes in geosynchronous orbits (e.g., \citealp{Kumar_et_al_2023}) and SuperMAG magnetometer network derived indices (SML and SMU) (e.g., \citealp{Fejer_et_al_2024}). We show Figure \ref{fig9} for this purpose, where we plot (i) electron fluxes at different energy channels as observed by GOES18-SEISS-MPSH and (ii) SML and SMU during 11-19 September 2023. The periods when $B_z$ $\approx$ 0, $>$ 0 and $<0$ as well as the intervals 1-5 are also marked in this figure similar to Figure \ref{fig1}. The color codes used in this Figure are also same as Figure \ref{fig1}. SML and SMU are basically perturbation in geomagnetic fields as measured from more than 100 magnetometers stationed around the world \citep{Newell_and_Gjerloev_2011b} and these indices reveal the auroral electrojet activity and thereby, the magnetospheric activity. On the other hand, dispersionless energetic electron/ion injection is observed during substorm in the Earth’s magnetotail \citep{Baker_and_Pulkkinen_1991, Reeves_and_Henderson_2001} if the satellite happens to be inside the injection front. From Figure \ref{fig9}, one can see intensification of auroral electrojet currents (particularly westward electrojet denoted by SML) and geosynchronous particle flux enhancements during intervals 2, 3 and 5 indicating presence of substorm activity in the magnetosphere.. 

\subsubsection{Intervals 2 and 3}
It is seen from Figure \ref{fig6} and Table \ref{tab1} that during intervals 2 and 3, spectral indices of ions observed by both ASPEX-STEPS units and SEISS-MPSH are $\approx$ 5. During these intervals energetic ions in the IP medium exhibited harder spectra with spectral indices 1.44 and 1.36, as recorded by EPAM-LEMS120. Panel (a) of Figure \ref{fig5} shows almost no enhancement in the ion fluxes was observed by EPAM-LEMS120 was there during intervals 2 and 3, ruling out presence of any SEP event. In fact, we do not understand the reasons for hard spectra at L1 during these intervals and no further effort is made to do so. A number of dispersionless enhancements in the GOES $H^+$ fluxes are observed in the vicinity of intervals 2 and 3. These flux enhancements do not seem to be related with the fluctuations in the ACE ion fluxes shown in panel (a). This  indicates  that energetic ions in the IP medium seem to have little effect on the magnetospheric energetic ion fluxes during intervals 2 and 3. This could be due to the  intermittent and relatively less southward (with mean IMF $B_z$ = -2.66 nT and -0.8 nT during intervals 2 and 3, respectively) IMF $B_z$ conditions (less efficient magnetic merging on the dayside). Softer ion spectra (with spectral indices $\approx$ 5) observed by AL1 and GOES during these intervals hint towards the dominance of substorm associated energetic ions in the magnetosphere during intervals 2 and 3. 

In addition, we see significant spectral softening (differences of 0.58 and 0.64) in ASPEX-STEPS-PS with respect to NP unit during interval 2 and 3, respectively. However, we do not investigate further the reason for this anisotropy in this work due to limited observations. 

\subsubsection{Interval 4c}
During interval 4c when IMF $B_z$ $>$ 0, spectral indices measured by ASPEX-STEPS units are 2.79 (PS) and 2.85 (NP) (see Table \ref{tab1} and Figure \ref{fig7}). Energetic ions in the IP medium show spectral index of 2.55. However, $H^+$ in the geosynchronous orbits (GOES) exhibit typical spectral index of 5. Although AL1 and GOES were inside the magnetosphere, significant differences between the spectral indices observed by ASPEX-STEPS and SEISS-MPSH during interval 4c are seen.  

As shown in Figure \ref{fig9}, there is no substorm activity during interval 4c. It is not immediately clear why GOES shows softer spectra with spectral index similar to substorm accelerated ions during this interval. It is known that energetic particle spectra in the plasma sheet follow spectral index of $\approx$ 5 \citep{Christon_et_al_1988} with or without substorms. It is noteworthy here that GOES was in the night side during interval 4c (see Figure \ref{fig2}). Whether spectral index of 5.05 observed by GOES in this case is associated with plasma sheet particles is difficult to ascertain in absence of any supporting measurement. Note, even under northward IMF B$_z$, dayside reconnection can happen in the magnetic field lines whose footprints are very close to poles and magnetic field can open up in a very narrow region and also, in a complex manner (for example, see \citealp{fuselier2024global}). Therefore, entry of energetic ions from the IP medium closer to magnetopause (where AL1 was located during 4c) and deep inside the magnetosphere (where GOES-18 was located) can vary significantly resulting in inhomogeneity in spatial distribution of energetic ions. This could be the case during interval 4c when ASPEX-STEPS units might have observed energetic ions dominated by the entry of decaying SEPs (see Figure \ref{fig5}), which reached close to the magnetopause through polar regions. This could have resulted into nearly 2.8 spectral indices of ions measured by ASPEX-STEPS. 

\subsubsection{Interval 5c}
As seen in Figure \ref{fig6} and Table \ref{tab1}, spectral index of energetic ions in the IP medium during interval 5c is 2.29 while that for magnetospheric $H^+$ is 3.09. ASPEX-STEPS units (PS and NP) recorded ion spectral index of 2.45 each. Note that during this interval, SEPs associated with ICME-3 shock dominated in the IP medium (see Figure \ref{fig5}) and IMF $B_z$ was negative. Further, the magnetosphere was populated by substorm associated (Figure \ref{fig9}) energetic particles during interval 5c. Therefore, there is a possibility of interplay of SEPs and substorm associated ions in the magnetosphere during this interval. AL1 being closer to the magnetopause (see Figure \ref{fig2}), ion spectra in ASPEX-STEPS units seem to be dominated by SEP ions, whereas, comparatively more dominance of substorm accelerated ions can be seen in GOES $H^+$ spectrum during this interval. 

\subsection{AL1 in the IP medium}
From Figure \ref{fig1} and \ref{fig2} it can be seen that AL1 was in the IP medium during intervals 4a and 5a. As can be seen in Figure \ref{fig5}, during these intervals, the main source of energetic ions in the IP medium was SEPs associated with shocks driven by ICME-2 and ICME-3.    

\subsubsection{Interval 4a}
Figure \ref{fig8} and Table \ref{tab1} show that the spectral indices observed by ASPEX-STEPS units during interval 4a are 2.65 (PS) and 2.67 (NP). EPAM-LEMS120 and SEISS-MPSH recorded spectral indices of 2.19 and 4.8, respectively, from the L1 point and in the geosynchronous orbits, respectively. It may be  noted that the differences in spectral indices between ASPEX-STEPS units and EPAM-LEMS120 are not within the error margins in interval 4a. Therefore, comparatively softer (than EPAM-LEMS120) ion spectra were observed by ASPEX-STEPS units during this interval. Although both AL1 (Figure \ref{fig1}) and ACE (Figure \ref{fig5}) observed decaying phase of SEPs associated with ICME-2, it is not clear why the AL1 spectra during 4a is softer than the spectrum observed by ACE. There are two basic differences in the observations of EPAM-LEMS120 and ASPEX-STEPS during these intervals: (1) locations of ACE and AL1 and (2) orientations of detector units. As can be seen from panel (iii) of Figure \ref{fig1}, the distance of AL1 from the Earth was $<$ $20 R_E$ during interval 4a. On the other hand, ACE was around the L1 point ($>$ $220 R_E$). ACE being a spin stabilized satellite, EPAM-LEMS120 integrates ions coming from a large annular cone (see \citealp{Gold_et_al_1998} for details). On the other hand, the look directions of ASPEX-STEPS-PS and NP units were fixed and are shown in Figure \ref{fig3}. Most probably, these two factors are responsible for dissimilar spectral indices observed by ASPEX-STEPS units and EPAM-LEMS120 during interval 4a.

Another important point to be noted that there was little impact of these SEPs into the magnetospheric particle fluxes as softer ($\approx 5$) $H^+$ spectrum is observed in the magnetosphere during interval 4a. Note that IMF $B_z$ was positive during this interval, which probably stopped a large number of SEPs in the IP medium to enter the magnetosphere.   

\subsubsection{Interval 5a}
During interval 5a, the spectral indices of ions observed by ASPEX-STEPS units and EPAM-LEMS120 are 1.47 (PS), 1.24 (NP), and 1.46, respectively. Interestingly, GOES $H^+$ spectrum became very harder with spectral index of 2.08. This time, we do not see much differences in the AL1 and ACE spectral indices. This suggests that during interval 5a, the SEP populations associated with ICME-3 shock were isotropically distributed between the locations of ACE and AL1. Whatever the values of spectral indices are, it is clear from Figure \ref{fig8} and Table \ref{tab1} that ion spectra in interval 5a are much harder. Even if substorm activities are seen in Figure \ref{fig9} during this interval, SEPs (with harder spectra), facilitated by negative IMF $B_z$, enter the magnetosphere from the IP medium and dominate in the $H^+$ spectrum as observed by SEISS-MPSH. 

\subsection{Additional evidence of entry of energetic ions from IP medium into the magnetosphere}
In order to check the connection between the energetic ions measured by ACE in the IP medium and GOES in the magnetosphere, we calculate correlation between ion fluxes in these two regions. As the energetic ions associated with these ICME shocks reach the Earth well before the arrival of the ICMEs, we choose intervals starting from 1.5 days (to keep uniformity) before (as shown in shaded regions with varying contrasts in Figure \ref{fig5}) the shock arrival times of these ICMEs for the correlation study. The average IMF $B_z$ values during the shaded intervals in Figure \ref{fig5} before the arrivals of shocks associated with ICME-1, 2, and 3 are -0.5, -0.49, and -2.6, respectively. This suggests that before the arrival of ICME-3 shock, IMF B$_z$ was predominantly southward, as compared to the intervals before the arrivals of ICME-1 and ICME-2 shocks. 

\begin{figure}[ht!]
\plotone{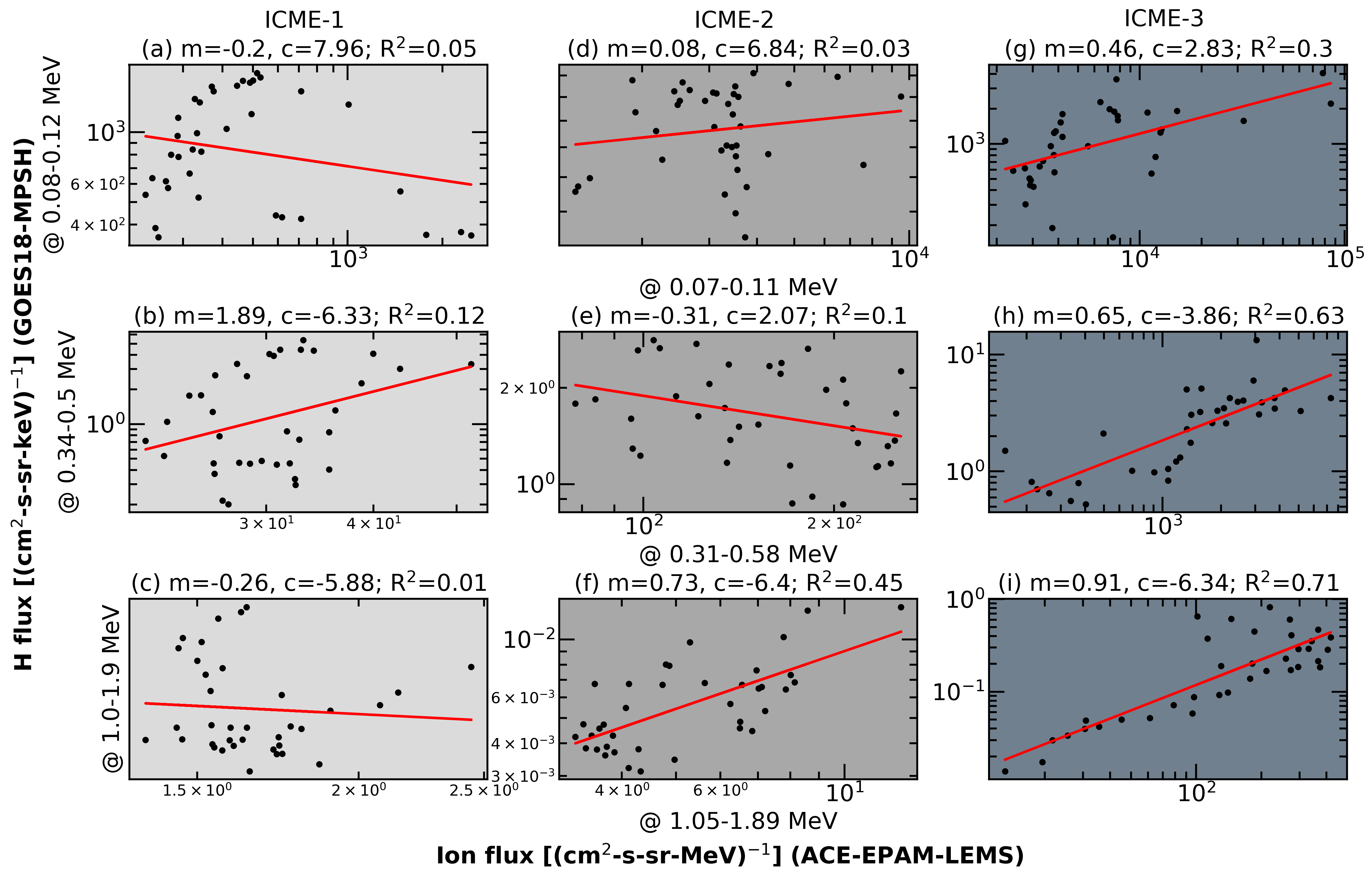}
\caption{Correlations between ion fluxes measured by ACE-EPAM-LEMS120 and proton fluxes measured by GOES18-SEISS-MPSH at three energy channels (three rows) corresponding to ICME-1 (first column), ICME-2 (middle column), and ICME-3 (last column). The red lines ($y = mx +c$) are the least-squared fit lines. 1 hour averaged fluxed have been used in this case. The corresponding fit parameters ($m$ and $c$) and $R^2$ values are mentioned at the top of each panel.  \label{fig10}}
\end{figure}

In Figure \ref{fig10}, we compare hourly averaged ion fluxes in almost three similar energy channels for EPAM-LEMS120 (0.07-0.11 MeV, 0.31-0.58 MeV, and 1.05-1.89 MeV) and SEISS-MPSH (0.08-0.12 MeV, 0.34-0.5 MeV, and 1.0-1.9 MeV). The slopes ($m$ values) of the best-fitted lines (red solid lines), intercepts ($c$ values), and square of the correlation coefficients ($R^2$) are mentioned at the top of each panel. It can be seen that $R^2$ is poor for ICME-1 (0.05, 0.12, and 0.01 for Panels a, b, and c respectively) and ICME-2 (0.03, 0.1, 0.45 for Panels d, e and f), but comparatively higher for ICME-3 (0.3, 0.63, 0.71 for Panels g. h, and i). It is also important to note the increase in $R^2$ values with increasing energy of ions in the second (panels d, e, and f) and third columns (panels g, h, and i) of Figure \ref{fig10}. This observation is particularly applicable for the third column (ICME-3). This suggests that predominantly negative IMF $B_z$ condition might have played an important role in allowing entry of energetic ions (generated by ICME-3 shock) into the magnetosphere. Further, we also note that during interval 5 (which is within the shaded interval before the ICME-3 shock), ion spectral index for ACE (see Table \ref{tab1}) changes from 1.46 to 2.29, which is a change by 0.83. The corresponding change for GOES is 1.01 (from 2.08 to 3.09). The changes in the spectral indices in ACE and GOES seem to be nearly commensurate here. Further, this change in GOES $H^+$ spectra is quite significant and the spectral indices are much harder ($\approx$ 2-3 in place of 5) with respect to the spectral indices obtained during the other intervals (i.e., 1, 2, 3, 4a, 4b, and 4c). Figure \ref{fig10} suggests that the hardening in the magnetospheric $H^+$ spectra probably occurred due to enhanced entry of higher energy ions generated from ICME-3 shock resulting in increasing  correlations between ACE and GOES fluxes with energy. This is not the case for the shaded interval before the ICME-2 shock, which includes interval 4. During interval 4, GOES spectral indices remained $\approx$ 5 (see Table \ref{tab1}) and the change (from 4.8 to 5.05) in spectral indices is not significant.

\subsection{On the determination of magnetopause and bow shock position}
The demarcation of magnetopause, magnetosheath and bow shock in the present work is based on model calculations and not by any collocated, in-situ measurements. Therefore, there can be certain degree of uncertainties associated with the determination of the locations of magnetopause, magnetosheath, and bow shock. The model calculation of magnetopause location in the present work is based on measured solar wind density and speed values. Since dynamic pressure (a function of solar wind density and speed) does not change significantly during interval 1-5, we do not expect significant change in the model derived boundary locations during these intervals. The work of \cite{Ingale_et_al_2019} indicates a variation of around $0.3 R_E$ in the model-derived magnetopause stand-off distance in the year 2020. On the other hand, the typical standard deviation expected for the bow shock nose stand-off distance is around $1.2 R_E$ \citep{Chao_et_al_2002}. Therefore, we feel that model-based determinations of different regions (magnetosphere, magnetosheath, and IP medium) will not significantly alter the conclusions drawn in this work.
 
\section{Conclusions}
Aditya-L1, India’s first space-based solar observatory, completed several Earth-bound orbits during 11 - 19 September 2023 before being inserted into the L1 halo orbit. During these transitions, the high-energy ion detectors (STEPS) were switched on to study the energetic ion fluxes of the magnetosphere, magnetosheath, and interplanetary space. Analyses of the ASPEX-STEPS data collected during the spacecraft's multiple transitions through these regions reveal magnetospheric energetic ion fluxes corresponding to the passage of multiple ICMEs and occurrence of substorms. Such data provide valuable insights into how the energetic ions from the interplanetary medium affect the magnetosheath and magnetosphere.

Based on the energetic proton spectra from 0.2 MeV to 2 MeV, and supported by data collected by other spacecraft such as ACE and GOES-18, we conclude that ASPEX-STEPS has observed softer ion spectra close to $\approx 5$ when magnetospheric substorms were present but no SEP event was present. It is suggested that the energetic ion fluxes in the magnetosheath is determined by the interplay of energetic ions reaching magnetosheath from the IP medium, bow shock, and magnetosphere. It is argued that the energetic ion fluxes inside the magnetosphere depends critically on the polarity of the IMF $B_z$ as it controls the penetration of energetic ions arriving from external drivers like ICME shock and competing with energetic ions coming from internal drivers like substorm. This competition eventually determines the spectral indices of the energetic ions inside the magnetosphere. Although occasional and inconsistent directional anisotropy (mild to significant) is seen in IP medium and magnetosphere, we see consistently mild directional anisotropy in the magnetosheath in three out of three cases. This is attributed to the spatial variation of the mixing of the energetic ions reaching magnetosheath from three different sources, i.e. the IP medium, bow shock, and magnetosphere. The measurements by ASPEX-STEPS also reveal that energetic ion spectra in the magnetosphere during the passage of two ICMEs vary significantly despite having similar geomagnetic impact. 

\section{Acknowledgements}
Authors also would like to thank various centers of ISRO for providing technical support and facilities for the test and calibration of ASPEX payload. Thanks are also due to project, mission, and review teams of ISRO for their support. Authors are also grateful to the principal investigators and members of the ACE-EPAM and GOES18-SEISS teams for generating and managing the datasets used in this work. This work is supported by the Department of Space, Government of India.

\section{Data availability statements}
The ASPEX-STEPS data used in this work are available at \url{https://zenodo.org/records/15708728}. Solar wind in situ parameters and ion fluxes measured by the ACE-EPAM-LEMS are taken from NASA GSFC CDAWeb (\url{https://cdaweb.gsfc.nasa.gov/index.html}). GOES-18 data are taken from NOAA website (\url{https://www.ngdc.noaa.gov/stp/satellite/goes-r.html}). SMU/SML data are available at \url{https://supermag.jhuapl.edu/indices/?fidelity=low&tab=description&layers=SME.UL}.
\bibliography{STEPS_paper}{}
\bibliographystyle{aasjournal}
\appendix

\section{Validation of ASPEX-STEPS data}
\begin{figure}[ht]
\centering
\includegraphics[width=0.75\textwidth, height=0.55\textwidth]{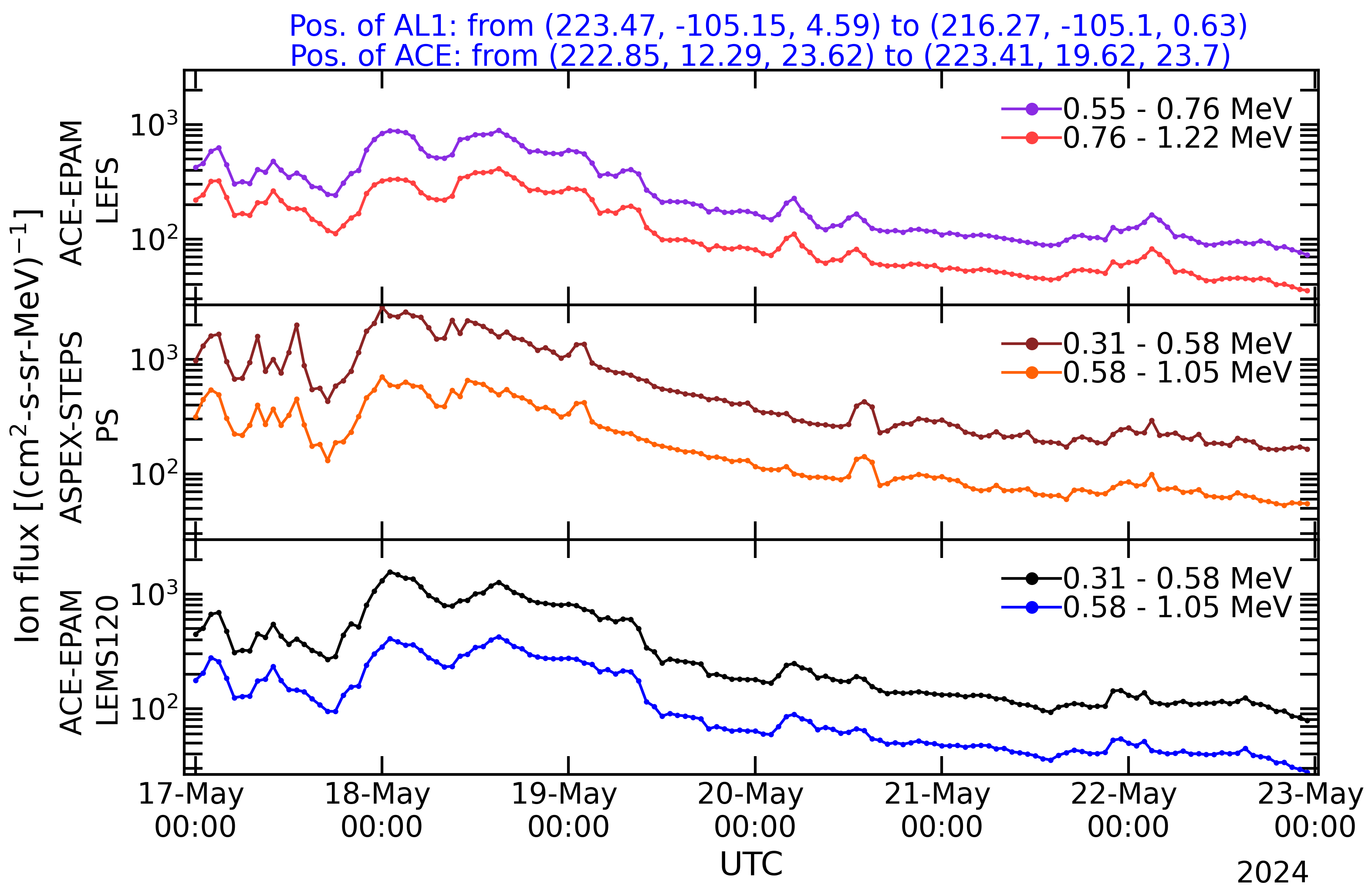}
\renewcommand{\thefigure}{A1}
\caption{Light curves of hourly averaged ion fluxes observed by ACE-EPAM-LEFS (top panel), ASPEX-STEPS-PS (middle panel), and ACE-EPAM-LEMS120 (bottom panel) at two energy channels during 17 – 22 May 2024 are shown. The positions of AL1 and ACE during this period in GSE coordinate system are mentioned at the top of the figure. Note that both AL1 and ACE were around the L1 point and are widely separated in Y-GSE direction during this interval. \label{fig:A1}}
\end{figure}

\begin{figure}[ht]
\centering
\includegraphics[width=0.8\textwidth, height=0.4\textwidth]{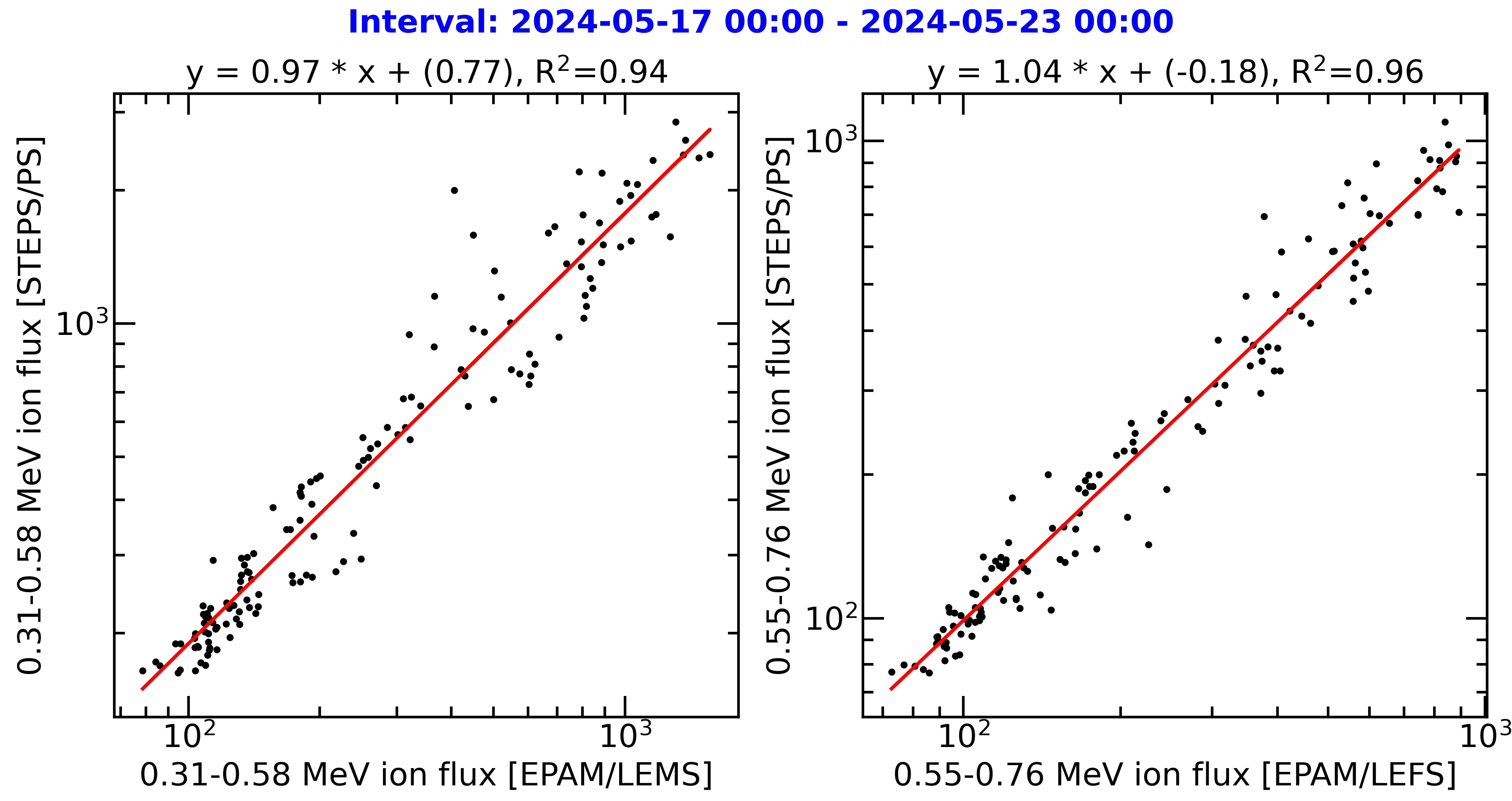}
\renewcommand{\thefigure}{A2}
\caption{Comparisons between hourly averaged ion fluxes measured by (left) ACE-EPAM-LEMS120 and ASPEX-STEPS-PS @ 0.31 – 0.58 MeV and (right) ACE-EPAM-LEFS and ASPEX-STEPS-PS @ 0.55 – 0.76 MeV for the interval 17 – 22 May 2024 (light curves are shown in Figure \ref{fig:A1}). The least square fit lines (with equation $y = mx +c$) are shown in red. The fitted equation and the square of the correlation coefficient ($R^2$) are mentioned at the top of each panel. It can be seen there is very good correlation between measurements of ASPEX-STEPS-PS and ACE-EPAM detectors.  \label{fig:A2}}
\end{figure}

\begin{figure}[ht]
\centering
\includegraphics[width=0.75\textwidth, height=0.55\textwidth]{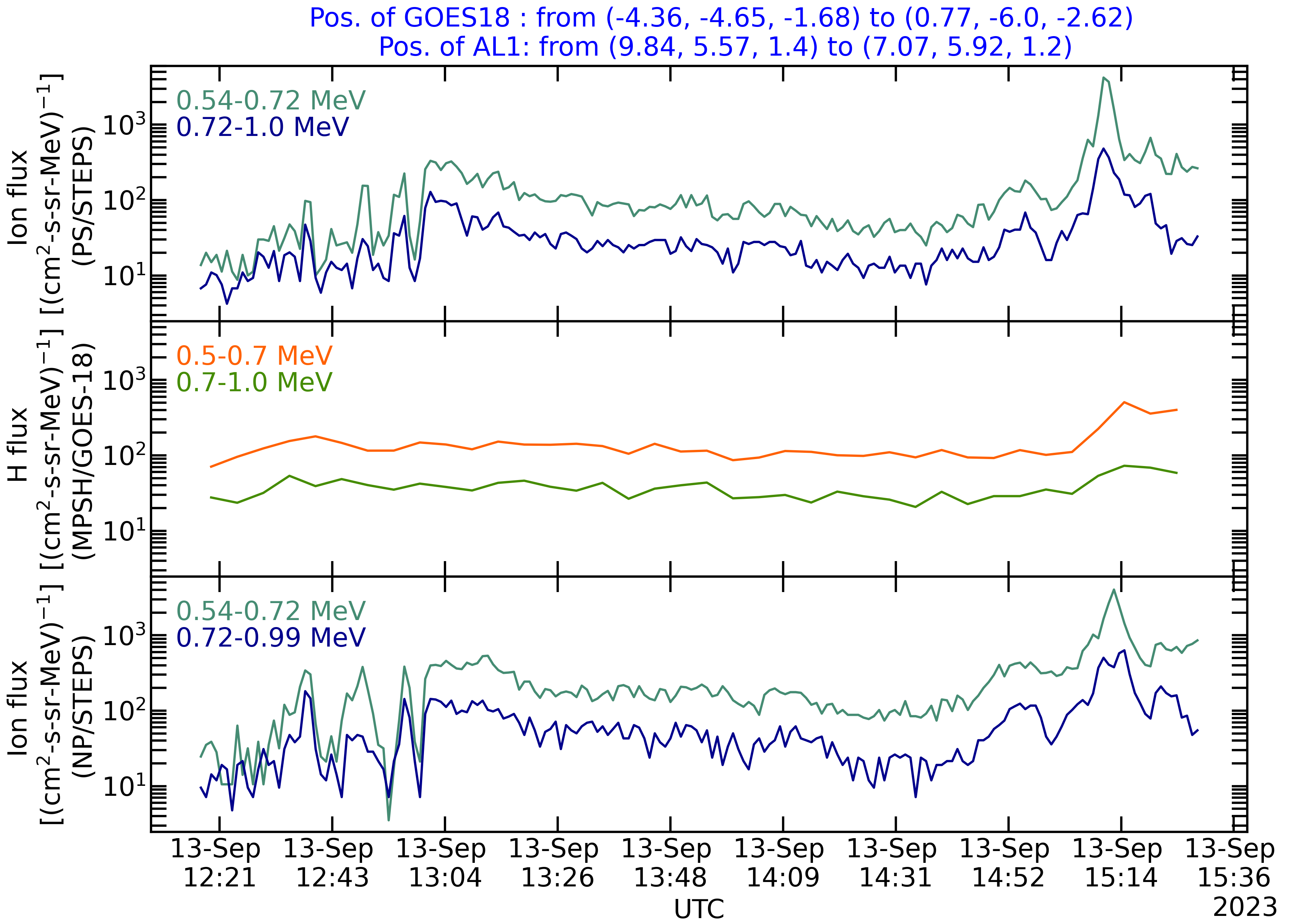}
\renewcommand{\thefigure}{A3}
\caption{Light curves of hourly averaged ion fluxes observed by ASPEX-STEPS-PS (top panel), proton fluxes observed by GOES18-SEISS-MPSH (middle panel), and ASPEX-STEPS-NP (bottom panel) at two energy channels (0.54 – 0.72 MeV and 0.72 – 1.0 MeV) during interval 2 (13 September 2023 12:18 UT – 15:31 UT) are shown. Note that the time resolution of STEPS fluxes is 1 minute and that of the GOES fluxes is 5 minutes. The positions of AL1 and GOES18 during this period in GSE coordinate system are mentioned at the top of the figure. \label{fig:A3}}
\end{figure}

\begin{figure}[ht]
\centering
\includegraphics[width=0.8\textwidth, height=0.4\textwidth]{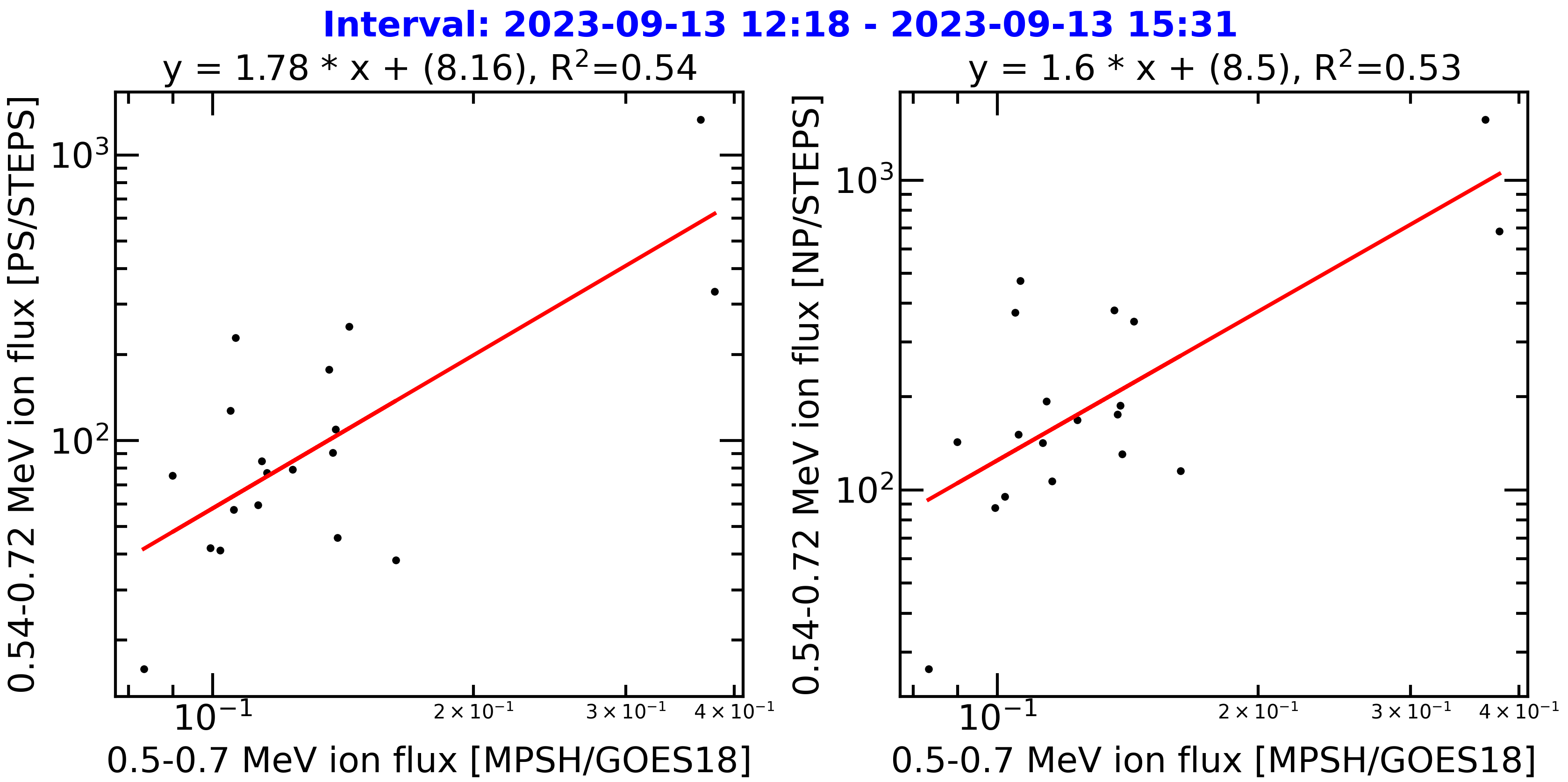}
\renewcommand{\thefigure}{A4}
\caption{Comparisons between 10 minutes averaged ion fluxes measured by (left) GOES18-SEISS-MPSH and ASPEX-STEPS-PS @ 0.54 – 0.72 MeV and (right) GOES18-SEISS-MPSH and ASPEX-STEPS-NP @ @ 0.54 – 0.72 MeV for the interval 13 September 2023 12:18 UT – 15:31 UT (light curves are shown in Figure \ref{fig:A3}) when AL1 was inside the magnetosphere. The least square fit lines (with equation $y = mx +c$) are shown in red. The fitted equation and the square of the correlation coefficient ($R^2$) are mentioned at the top of each panel. It can be seen there is reasonable correlation between measurements of ASPEX-STEPS-PS and GOES18-SEISS-MPSH detectors. \label{fig:A4}}
\end{figure}

In order to validate ASPEX-STEPS data with respect to data obtained from existing satellites like ACE (when AL1 was in the IP medium) and GOES18 (when AL1 was in the magnetosphere), a comparison study is presented here. Figure \ref{fig:A1} and \ref{fig:A2} illustrate the comparison of ion fluxes observed by instruments (EPAM-LEMS120 and LEFS) on board ACE with those observed by ASPEX-STEPS when AL1 was in its nominal configuration (at L1) in the IP medium. We compare 0.31-0.58 MeV (0.55-0.76 MeV) ion fluxes observed by ASPEX-STEPS-PS and EPAM-LEMS120 (EPAM-LEFS) (see Figure \ref{fig:A2}). On the other hand, Figure \ref{fig:A3} and \ref{fig:A4} provide the comparison of proton fluxes observed by GOES18-SEISS-MPSH with ASPEX-STEPS detector units. In this case, ion fluxes in the energy range of 0.54 – 0.72 MeV as observed by ASPEX-STEPS-PS and NP detector units are compared with proton fluxes observed by GOES18-SEISS-MPSH in the similar energy range (see Figure \ref{fig:A4}). 
\section{Spectral index of substorm associated protons}
\begin{figure}[ht!]
\plotone{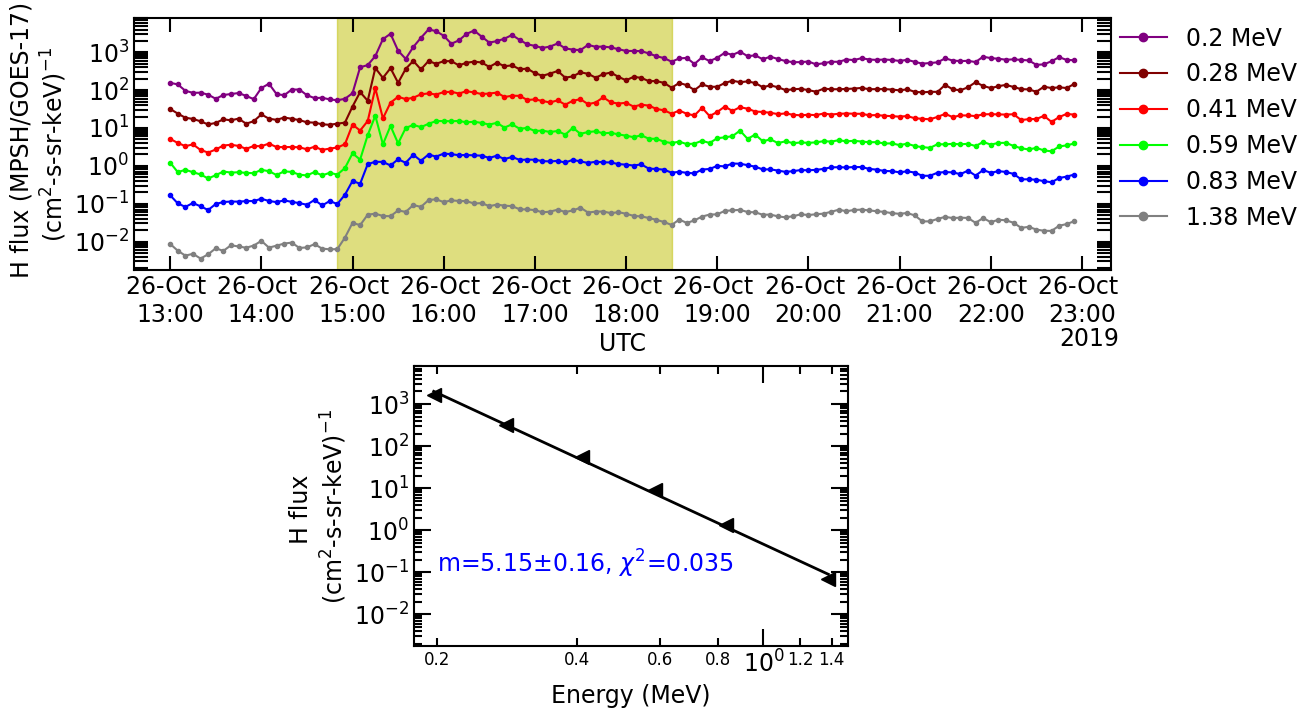}
\renewcommand{\thefigure}{A5}
\caption{Top panel: temporal variations of proton fluxes at different energy channels as observed by GOES17-SEISS-MPSH during 26 October 2019 13:00 UT – 17:00 UT. \cite{Rathi_et_al_2025} reported two substorm events during this interval. The yellow shaded interval corresponds to one of those substorms. Bottom panel: proton spectrum averaged over the yellow shaded interval shown above. Spectral index ($m$) and goodness ($\chi ^2$ value) of the spectral fit (black solid line) are mentioned in blue. It is to be noted that the spectral index of protons associated with magnetospheric substorm is $\geq$ 5. \label{fig:A5}}
\end{figure}

In order to check the typical spectral index of protons associated with magnetospheric substorm in the magnetosphere, we select a previously reported substorm event from \cite{Rathi_et_al_2025} and calculate proton spectral index during this event. The event spans the interval of 26 October 2019 13:00 UT – 17:00 UT. Figure \ref{fig:A5} shows the temporal variations of protons at different energy channels as measured by GOES17-SEISS-MPSH during the substorm event  and correspondong spectral index. This indicates that the spectral index of substorm-associated protons in the geosynchronous orbit is $\geq$ 5. We have tested a few more substorms (on 10 January 2025, 11 January 2025, 17 January 2025, 14 February 2025, 18 March 2025, and 19 March 2025), which are not shown here, and the spectral indices are close to 5 as well.

\end{document}